\documentclass[aps,prb,twocolumn,amsfonts,showpacs,longbibliography,superscriptaddress]{revtex4-2}
\RequirePackage{snapshot}
\usepackage{verbatim,amsmath,amssymb,bm,graphicx,bbold,amsthm,amsfonts}
\usepackage{standalone}
\usepackage[toc,page,titletoc]{appendix}
\usepackage[bottom]{footmisc}
\usepackage{hyperref}
\hypersetup{
    colorlinks=true,
    linkcolor=blue,
    citecolor=blue,
    filecolor=blue,
    urlcolor=blue,
}

\usepackage[capitalize]{cleveref} 
\usepackage{enumitem}
\usepackage{soul}
\usepackage{framed}
\usepackage{mathrsfs}
\usepackage{esint}
\setlist{nosep}
\usepackage{placeins}
\usepackage[all]{xy}
\usepackage{color}
\usepackage[utf8]{inputenc}
\usepackage{float}
\usepackage{natbib}
\usepackage{tikz-cd}
\usepackage{leftidx}
\usepackage[normalem]{ulem}
\usepackage[percent]{overpic}

\newcommand{\ket}[1]{\mbox{$ | #1 \rangle $}}
\newcommand{\bra}[1]{\mbox{$ \langle #1 | $}}
\newcommand{\braket}[2]{\mbox{$ \langle #1 | #2 \rangle $}}
\newcommand{\abs}[1]{\left|#1\right|}

\newcommand\beq {\begin{equation}}
	\newcommand\eeq {\end{equation}}
\newcommand\beqa {\begin{equatiobn}\begin{array}}
		\newcommand\eeqa {\end{array}\end{equation}}
\newcommand\bal {\begin{align}}
	\newcommand\eal {\end{align}}
\newcommand{\bea}{\begin{eqnarray}}
	\newcommand{\eea}{\end{eqnarray}}

\newcommand{\innerproduct}[2]{\langle #1| #2\rangle}

\begin{document}

\title{Continuum Fractons: Quantization and the Few Body Problem}

\author{Ylias Sadki}
\email{ylias.sadki@physics.ox.ac.uk}
\affiliation{Rudolf Peierls Centre for Theoretical Physics, University of Oxford, Oxford OX1 3PU, United Kingdom}
\author{Abhishodh Prakash}
\email{abhishodhprakash@hri.res.in} 
\altaffiliation{(he/him/his)}
\affiliation{Rudolf Peierls Centre for Theoretical Physics, University of Oxford, Oxford OX1 3PU, United Kingdom}
\affiliation{Harish-Chandra Research Institute,  a CI of Homi Bhabha National Institute, Prayagraj (Allahabad) 211019, India}
\author{S. L. Sondhi}
\email{shivaji.sondhi@physics.ox.ac.uk}
\affiliation{Rudolf Peierls Centre for Theoretical Physics, University of Oxford, Oxford OX1 3PU, United Kingdom}
\begin{abstract}
We formulate a continuum quantum mechanics for non-relativistic, dipole-conserving fractons.
Imposing symmetries and locality results in novel phenomena absent in ordinary quantum mechanical systems. 
A single fracton has a vanishing Hamiltonian, and thus its spectrum is entirely composed of zero modes. 
For the two-body problem, the Hamiltonian is perfectly described by Sturm--Liouville (SL) theory. 
The effective two-body Hamiltonian is an SL operator on $(-1,1)$ whose spectral type is set by the edge behavior of the pair inertia function $K(x)\sim \lvert x -x_\mathrm{edge} \rvert^{\theta}$. 
We identify a sharp transition at $\theta=2$: for $\theta<2$ the spectrum is discrete and wavepackets reflect from the edges, whereas for $\theta>2$ the spectrum is continuous and wavepackets slow down and, dominantly, squeeze into asymptotically narrow regions at the edges.
For three particles, the differential operator corresponding to the Hamiltonian is piecewise defined, requiring several `matching conditions' which cannot be analyzed as easily. 
We proceed with a lattice regularization that preserves dipole conservation, and implicitly selects a particular continuum Hamiltonian that we analyze numerically. 
We find a spectral transition in the three-body spectrum, and find evidence for quantum analogs of fracton attractors in both eigenstates and in the time evolution of wavepackets. We provide intuition for these results which suggests that the lack of ergodicity of classical continuum fractons will survive their quantization for large systems.
\end{abstract}

\maketitle
\tableofcontents

\section{Introduction}
The study of `fractons'~\cite{NandkishoreHermeleFractonsannurev-conmatphys-031218-013604,PretkoChenYou_2020fracton,GromovRadzihovsky2022fractonReview} i.e., particles with restricted or no mobility, has gained much prominence in recent years.
Although the original motivations for their study can be traced to attempts at understanding non-trivial topological order~\cite{Chamon_Fracton_PhysRevLett.94.040402,Haah_FractonPhysRevA.83.042330,VijayHaahFu_FractonDuality_PhysRevB.94.235157}, the study of `fundamental' fractons conserving multipole moments has been enormously fruitful and has yielded insights by challenging conventional wisdom of universality, emergence and ergodicity~\cite{GorantlaLamSeiberg_UVIR_PhysRevB.104.235116,YouMoessner_UVIR_PhysRevB.106.115145,KhemaniHermeleNandkishore_Shattering_PhysRevB.101.174204,SalaRakovskyVerresenKnapPollmann_FragmentationPhysRevX.10.011047}. 

Most of the earlier focus has been on lattice quantum systems, and their effective field theories~\cite{Lake2022DipolarBH,Lake2022NonFermi,Yuan2020FractonicSF,Chen2021FractonicSF2}.
In a recent series of works~\cite{AP2023NRFractons,AP2024MachianFractons,babbar2025classicalfractonslocalchaos,sadki2025phasespacefractons}, we initiated and explored classical fractons in the continuum as a natural extension of ordinary classical Hamiltonian dynamics in the presence of global multipole conservation symmetry.

We showed that classical fractons have several noteworthy features in their few and many-body dynamics: 
i) they exhibit `Machian' dynamics --- a term originally introduced for fractons in \cite{pretko2017emergent} --- wherein isolated particles are immobile and motion occurs only in the presence of multiple particles in close proximity~\cite{AP2023NRFractons}.
ii) their dynamics is characterized by attractors in configuration (position-velocity) space, seemingly violating Liouville's theorem~\cite{AP2023NRFractons}.
iii) the attractors result in strong breaking of ergodicity and the appearance of non-equilibrium steady states that cannot be described by conventional (Gibbsian) statistical ensembles~\cite{AP2024MachianFractons}.
iv) the attractors and steady states manifest themselves by particles breaking translation symmetry by clustering in arbitrary dimensions and energy density, evading the theorem of Hohenberg, Mermin, Wagner and Coleman~\cite{AP2024MachianFractons}, 
v) global ergodicity breaking is accompanied by local chaos within clusters and the generic emergence of a bi-directional arrow of time~\cite{babbar2025classicalfractonslocalchaos}.  
We also showed that dipole conservation symmetry is important to preserve all these features~\cite{AP2023NRFractons} which disappear with explicit symmetry breaking.
These dipoles generalize to conservation of multipole moments in position~\cite{AP2024MachianFractons} and momentum space~\cite{sadki2025phasespacefractons} .

With the classical picture in hand, the present paper addresses the following general questions. 
Upon canonical quantization, what are the spectrum and eigenfunctions of the Hamiltonian? 
How does the dynamical evolution proceed, and does the quantized system exhibit the attractors and the broken ergodicity seen classically? 
Our results are organized by particle number $n$: for $n=2$ we obtain an essentially complete solution; for $n=3$ we give partial results based primarily on computation. 
Throughout, we treat distinguishable particles, noting that extending the formalism to bosons or fermions is straightforward: the clustering properties are the same for identical, fermionic, or bosonic particles.
The excitations we study are strict fractons: particles that are strictly non-interacting beyond a finite separation, as in standard lattice formulations. 
A new mathematical feature is that the classical Hamiltonian possesses internal points/lines in phase space; for $n=2$ these become boundary conditions that permit the use of standard theory on intervals, whereas for $n\ge 3$ they raise further mathematical questions. 
Readers coming from lattice fracton models may wish to begin with Appendix~\ref{sec:lattice_models}, where we spell out the detailed connection between our continuum formulation and lattice models.

The rest of the paper is organized as follows: after reviewing the classical framework and setting up canonical quantization (including a discussion on two sources of zero modes) in \cref{sec:recap}, we analyze the two-fracton problem: in \cref{sec:spectrum_transition} we identify a discrete-to-continuous spectral transition controlled by the edge behavior of $K(x)$; in \cref{sec:discrete_spectrum} we formulate the self-adjoint extensions in the limit-circle case and exactly solve a case of the Hamiltonian; in the complementary limit-point regime we develop a semiclassical/WKB construction showing that wavepackets do not reflect but instead pile up at the edges, a picture we develop using a Liouville transform (\cref{sec:liouville_transform}). 
We then turn to three fractons in \cref{sec:three_fractons}, presenting numerics that (i) low-energy states localize along the classical attractor lines while allowing tunneling between sectors related by permutation and (ii) exhibit a finite-size crossover in level spacings consistent with a discrete-to-continuum transition whose apparent threshold drifts toward $\theta_c\!\approx\!2$. 
In \cref{sec:comment_on_many_fracton}, we comment on the relation to Hilbert Space Fragmentation in lattice models, and links to continuum field theory fracton models.
Appendix~\ref{sec:lattice_models} details a lattice discretization and its induced boundary conditions, and Appendix~\ref{sec:three_fracton_internal_lines} analyzes boundary/matching conditions on internal $K=0$ lines.
\cref{app:hexagon_symmetries} looks at the Dihedral symmetry present in the three body problem, and its implications on the quantum spectrum, and then the generalization to the symmetries of the $N$ fracton problem, and how to restrict to fermions and bosons.
\cref{app:continuum} further details the link of our fractonic model to field theory models.

\section{Recap of classical fractons and quantum preliminaries}
\label{sec:recap}
We begin with a brief review of the formulation of classical dynamics of fractons~\cite{AP2023NRFractons,AP2024MachianFractons,babbar2025classicalfractonslocalchaos,sadki2025phasespacefractons}. 
We will work within the Hamiltonian framework where the configuration of the system is specified by $Nd$ positions $x^\mu_i$ and momenta $p^\mu_i$ where $\mu=1,\ldots, d$ denotes the spatial dimension and $i=1,\ldots,N$ labels the particles. 
From now on, we will only consider the case of $d=1$. 
Imposing the conservation of the total dipole moment $D = \sum_{i=1}^{N}x_i$ corresponds to the invariance of the Hamiltonian under the shifts of each momentum coordinate $p_i \rightarrow p_i + a$. 
We consider a system of identical particles with translation invariance in the position space ($x_i \mapsto x_i + b$) and demand the locality and boundedness of the energy from below. 
This results in the following form of the Hamiltonian.

\begin{equation}
H = \sum_{i<j=1}^N \left[ \frac{\left({p}_i - {p}_j \right)^2}{2} K(\abs{{x}_i - {x}_j}) + U(\abs{{x}_i - {x}_j}) \right] \label{eq:H_dipole}
\end{equation}

The function $K(\dots)$ has local support and is strictly non-negative. 
$U(\dots)$ is a two-body interaction that is also assumed to be local. 
As done previously, we will choose to set $U=0$ and focus on the kinetic term. 
It is immediately clear that the equations of motion for a single particle $N=1$ are trivial: $\dot{x} = \dot{p} =0$. 
The unusual disassociation of momenta with velocity that is seen for a single particle also continues for an arbitrary number as seen through Hamilton's equation of motion
\begin{equation}
    \dot{x}_i = \sum_{j \neq i} (p_i - p_j) K(\abs{{x}_i - {x}_j}). \label{eq:velocity_momenta}
\end{equation}
This tells us that the velocity of particle $i$ is determined by the difference of its momentum with those of all proximate particles, i.e. for which $K(\abs{{x}_i - {x}_j})$ is non-vanishing. 

\begin{figure}[htbp]
\centerline{\includegraphics[width=8.6cm]{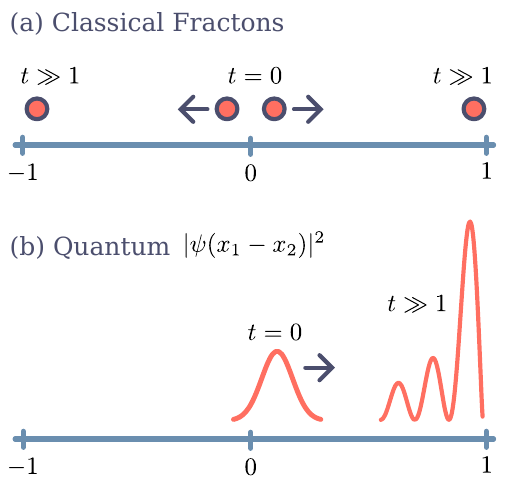}}
\caption[]{The classical-quantum correspondence of trajectories for 2 particles. \emph{(a)}: A classical trajectory for two particle, with $H = \frac{1}{2}K(x_1-x_2)(p_1-p_2)^2$ generically spreads out to $\abs{x_1 - x_2} = \pm 1$. \emph{(b)}: The analogous quantum trajectory: two particles starting off in proximity of eachother, with a localized wavepacket in reduced coordinates $x_1 - x_2$. Remarkably, for $K(x)$ vanishing as $K \sim (1 - \abs{x})^\theta$ with $\theta > 2$, the quantum fractons behave classically: at late times, the wavefunction amplitude piles up at the edge: i.e. the particles are localized with a separation of $x_1 - x_2 = 1$. We explore classical-quantum correspondence in this paper, to explore how the classical clustering generalizes to the quantized problem.  \label{fig:quantum_sketch} } 
\end{figure}

Looking closer at the Hamiltonian, it is constructed of pairs of $(p_i - p_j)^2 K(\lvert x_i - x_j \rvert)$ terms.
While both the total momentum $P = \sum_i p_i$ and the energy $H$ must remain constant, individual momenta $p_i$ and $p_j$ may grow off to positive or negative infinity, as the increasing $(p_i - p_j)^2$ terms are compensated by the decaying $K(\lvert x_i - x_j \rvert)$ terms, which ensure $P$ and $H$ remain constant.
Hence, individual momenta are unbounded at finite energy density, which implies an unbounded phase space. 
This leads to non-ergodicity, as an unbounded phase space cannot be explored in an ergodic fashion, leading to the formation of clustered states.
A cartoon two particle clustering is shown in \cref{fig:quantum_sketch}(a).

Finally, it was shown that the conservation of total dipole moment and momentum motivates us to change variables through a canonical transformation $\{x_j,p_j\} \mapsto \{q_\alpha,\pi_\alpha\}$:
    \begin{align}
        q_{\alpha = 1,\ldots,N-1} &= \frac{\sum_{j=1}^\alpha x_j - \alpha x_{\alpha + 1}}{\sqrt{\alpha \left(\alpha + 1\right)}}, ~q_N = \frac{\sum_{j=1}^N x_j}{\sqrt{N}}, \nonumber\\
        \pi_{\alpha = 1,\ldots,N-1} &= \frac{\sum_{j=1}^\alpha p_j - \alpha p_{\alpha + 1}}{\sqrt{\alpha \left(\alpha + 1\right)}}, ~\pi_N = \frac{\sum_{j=1}^N p_j}{\sqrt{N}}. \label{eq:reduced_coordinates}
    \end{align}
Notice that the canonical pair $q_N \propto \sum_j x_j$ and $\pi_N \propto \sum_j p_j$ are  conserved quantities, being the total dipole moment and total momenta; thus, $q_N$ and $\pi_N$ are both constants, disappearing from the Hamiltonian and reducing the classical Hamiltonian from $2N$ degrees of freedom to $2N-2$.
The transformation is linear in coordinates, so in the quantum mechanical problem, the transform is unitary, and reduces the degrees of freedom by one.

\subsection{Canonical Quantization}
We now proceed to canonically quantize our system of classical fractons. We elevate the phase space coordinates to operators 
\begin{equation}
    x_j \mapsto \hat{x}_j, ~p_j \mapsto \hat{p}_j \label{eq:canonical quantization}
\end{equation}
and impose the canonical commutation relation (setting $\hbar =1$), 
\begin{equation}
    [\hat{x}_i,\hat{p}_j] = i \delta_{ij}.
\end{equation}
An operator ordering ambiguity presents itself---the kinetic term of \cref{eq:H_dipole} can be written in one of two ways
\begin{multline}
    {\left({p}_i - {p}_j \right)^2} K(\abs{{x}_i - {x}_j}) \rightarrow \frac{\left(\hat{p}_i - \hat{p}_j \right)^2}{2} K(\abs{\hat{x}_i - \hat{x}_j}) +  h.c.\\
    \text{ or } 
     \left(\hat{p}_i - \hat{p}_j \right) K(\abs{\hat{x}_i - \hat{x}_j}) \left(\hat{p}_i - \hat{p}_j \right).\label{eq:ordering ambiguity}
\end{multline}

In this work, we will not explore the consequences of this ambiguity but choose the second ordering in \cref{eq:ordering ambiguity}. In the position basis, the Hamiltonian we will consider is

\begin{equation}
    \hat{H} = -  \sum_{i<j=1}^N    \left(\frac{\partial}{\partial x_i} - \frac{\partial}{\partial x_j} \right) \frac{K(\abs{{x}_i - {x}_j})}{2} \left(\frac{\partial}{\partial x_i} - \frac{\partial}{\partial x_j} \right).\label{eq:H_quantum_general}
\end{equation}

\subsection{Two sources of zero modes and degeneracy}
\label{sec:zeromode}
\begin{figure}
    \centering
    \begin{tikzpicture}[scale=.5]
\definecolor{colour2}{HTML}{ff6f61}
\definecolor{colour1}{HTML}{6a8eaf}
\draw[thick,colour2] (-5,0) -- (-1,0);
\draw[thick,colour2] (5,0) -- (1,0);
\node[] at (5.55,0) {$q_1$};
\node[] at (0,-1) {(a) Two particles};
\end{tikzpicture}
\begin{tikzpicture}[scale=.5]
\definecolor{colour2}{HTML}{ff6f61}
\definecolor{colour1}{HTML}{6a8eaf}
\definecolor{colour3}{HTML}{4b4e6d}

\node[] at (5.55,-4.9) {$q_2$};
\node[] at (-5.,5.3) {$q_1$};
\fill[colour1,opacity=0.5] (2.85/1.73,4.85) -- (0,2) -- (-2.85/1.73,4.85) --cycle;
\fill[colour1,opacity=0.5] (5,1) -- (1.73,1) -- (3.95,1.73*3.95-2) -- (5,1.73*3.95-2) --cycle;
\fill[colour1,opacity=0.5] (-5,1) -- (-1.73,1) -- (-3.95,1.73*3.95-2) -- (-5,1.73*3.95-2) --cycle;
\fill[colour1,opacity=0.5,rotate=180] (-5,1) -- (-1.73,1) -- (-3.95,1.73*3.95-2) -- (-5,1.73*3.95-2) --cycle;
\fill[colour1,opacity=0.5,rotate =180] (5,1) -- (1.73,1) -- (3.95,1.73*3.95-2) -- (5,1.73*3.95-2) --cycle;
\fill[colour1,opacity=0.5,rotate=180] (2.9/1.73,4.87) -- (0,2) -- (-2.9/1.73,4.87) --cycle;


\draw[colour2,opacity=1, thick] (-5,1) -- (5,1);
\draw[colour2,opacity=1,thick] (-5,-1) -- (5,-1);
\draw[colour2,opacity=.5,thick] (-1.75,-1) -- (-1.75,1);
\draw[rotate=60,colour2,opacity=.5,thick] (-1.75,-1) -- (-1.75,1);
\draw[rotate=120,colour2,opacity=.5,thick] (-1.75,-1) -- (-1.75,1);
\draw[rotate=-60,colour2,opacity=.5,thick] (-1.75,-1) -- (-1.75,1);
\draw[rotate=-120,colour2,opacity=.5,thick] (-1.75,-1) -- (-1.75,1);
\draw[rotate=180,colour2,opacity=.5,thick] (-1.75,-1) -- (-1.75,1);
\draw[rotate = 60,colour2,opacity=1,thick] (-6.1,1) -- (5,1);
\draw[rotate = 60,colour2,opacity=1,thick] (-5,-1) -- (6.1,-1);
\draw[rotate = -60,colour2,opacity=1,thick] (-5,1) -- (6.1,1);
\draw[rotate = -60,colour2,opacity=1,thick] (-6.1,-1) -- (5,-1);

\draw[->,thick,colour3] (-5,-4.85) -- (5.2,-4.85);
\draw[->,thick,colour3] (-5,-4.85) -- (-5,5);
\node[] at (0,-6) {(b) Three particles};
\end{tikzpicture}
    \caption{The reduced coordinates for (a) two particles and (b) three particles. Any square-integrable wave function with support in the blue shaded region corresponds to an exact zero-energy eigenstate of the Hamiltonian, with isolated fractons. This region is unbounded. Finite energy eigenstates are supported in the unshaded regions. }
    \label{fig:zero_mode}
\end{figure}
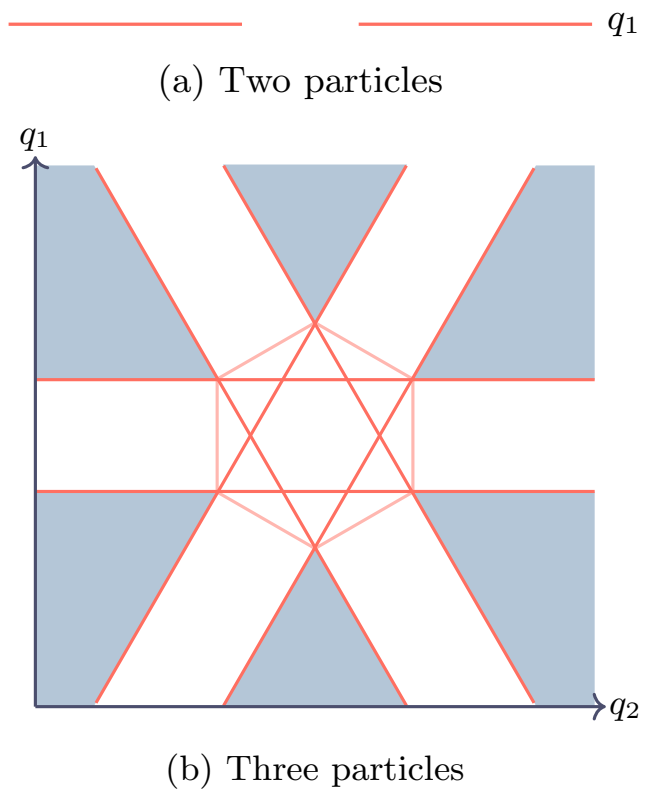
The Hamiltonian for a single particle is $H=0$ and the entire spectrum consists of an infinite number of zero modes. This seemingly trivial statement surprisingly generalizes to any number of particles. To see this, we impose the classical canonical transformation in \cref{eq:reduced_coordinates} through a unitary change of basis to simplify \cref{eq:H_quantum_general} as follows
\begin{multline}
    W \hat{H}(\hat{x}_1,\ldots,\hat{x}_N,\hat{p}_1,\ldots,\hat{p}_N) W^\dagger \\= \hat{H}(\hat{q}_1,\ldots,\hat{q}_{N-1},\hat{\pi}_1,\ldots,\hat{\pi}_{N-1}). 
\end{multline}
In this form, the Hamiltonian does not contain, and thereby trivially commutes with, the operators $\hat{q}_N$ and $\hat{\pi}_N$~\footnote{In the classical case, one can simply fix the center-of-mass position $q_N$ and momentum $\pi_N$ to definite values. In the quantum case, however, the operators $\hat{q}_N$ and $\hat{\pi}_N$ do not commute, satisfying $[\hat{q}_N, \hat{\pi}_N] = i$. Their values are therefore subject to the uncertainty principle and cannot be simultaneously fixed to arbitrary precision.}.
Consequently, these conserved operators satisfy the Heisenberg algebra
\begin{equation}
    [\hat{q}_N,\hat{\pi}_N] = i, \label{eq:Heisenberg}
\end{equation}
and hence generate an infinite dimensional zero-energy space, leading to an infinite degeneracy of each energy level. This large degeneracy, which arises from the non-Abelian nature of the conservation laws, can be accounted for by working in a reduced Hilbert space that is an eigenspace of the conserved quantities, effectively ignoring $\hat{q}_N, \hat{\pi}_N$ and working with the Hamiltonian operating on $N-1$ degrees of freedom. However, here too there can exist a large number of zero modes. To see this, consider a system of two particles whose Hamiltonian is written in reduced coordinates
\begin{equation}
    \hat{H} = \hat{\pi}_1 \ K(\sqrt{2} \hat{q}_1) \  \hat{\pi}_1 =- \frac{d }{d q_1} K(\sqrt{2} {q}_1)  \frac{d }{d q_1}. \label{eq:2-fracton_reduced}
\end{equation}
If we study the system on the real line $x_i \in \mathbb{R}$ and if we consider $K(\dots)$ to have strictly compact support in some domain, we see that outside this domain, the Hamiltonian vanishes. 
Thus, any square-integrable function outside this domain is an exact eigenstate with zero energy. 
Thus, the system has another source of infinite zero modes that generates a large spectral degeneracy. 
This generalizes to any finite number of particles so long as we have a compact $K(\dots)$ and the positions of the particles can extend over the full real line. 
In \cref{fig:zero_mode}, we show the region in reduced coordinates for two and three particles where we find zero-modes aside from those arising from the conserved charge operators. 
In what follows, we only focus on the finite energy spectrum, ignoring all exact zero modes.

\section{Two fractons}

\subsection{Spectral transition in the two-fracton problem}
\label{sec:spectrum_transition}
\subsubsection{Classes of two-fracton spectra}
Let us now consider the two-fracton Hamiltonian \cref{eq:2-fracton_reduced} written in a simplified form
\begin{equation}
    \hat{H} =    - \frac{d}{dx} K(x) \frac{d}{dx}. \label{eq:2_reduced_simplified}
\end{equation}
We will also assume that $K(x)$ has support only in the range $x \in (-1,1)$ and vanishes outside of it. 
As argued in \cref{sec:zeromode}, the operator has a large set of zero-modes localized outside the support of $K(x)$. 
The zero modes form the dominant part of the spectrum.
Here we will focus on the non-zero energy states that live within $x \in (-1,1)$. 
$\hat{H}$ is a Sturm-Liouville (SL) operator defined in a finite domain but not of the Schr\"{o}dinger type which describes conventional (non-fractonic) quantum mechanics. 
We highlight this to make a connection with the vast literature on the study of spectral properties of SL operators \cite{zettl2005sturm,littlejohn2011legendre}. 
General SL operators are of the form $\frac{d}{dx} p(x) \frac{d}{dx} - q(x)$, and have features that are absent in the more restrictive Schr\"{o}dinger operators. 
For instance, they can undergo a spectral transition from a discrete to a continuous spectrum. 
As shown in Ref.~\cite{stuart2018stability}, the SL operator in  \cref{eq:2_reduced_simplified} can have a discrete spectrum iff there exists $0 < \alpha < 2$ such that the following limit is non-zero:
\begin{equation}
\lim_{x \rightarrow \pm 1} \frac{K(x)}{(1-x^{2})^{\alpha}}. \label{eq:discrete_condition}
\end{equation}
\begin{figure}[!h]
    \centering
    \includegraphics[width=8.6cm]{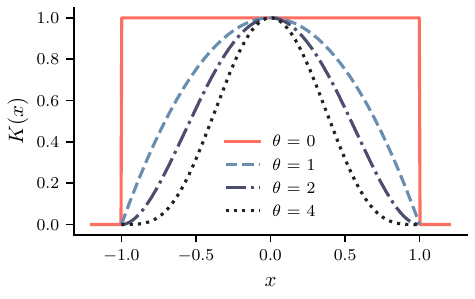}
    \caption{\cref{eq:K_theta} for various values of $\theta$. The spectral transition in the two-fracton problem occurs at $\theta = 2$ with $\theta<2$ hosting a discrete spectrum and $\theta>2$ hosting a continuous spectrum.}
    \label{fig:K_theta}
\end{figure}
The condition in \cref{eq:discrete_condition} can be translated into one on how the pair inertia function $K(x)$ vanishes at its edges $x = \pm 1$. If $K(x)$ vanishes as $(1-|x|)^\theta$, we see from \cref{eq:discrete_condition} that for $\theta<2$, we get a discrete spectrum and for $\theta>2$, we get a continuous one with a spectral transition at $\theta = 2$. We will demonstrate this using a semiclassical argument as well as mapping the SL operator in \cref{eq:2_reduced_simplified} to an equivalent Schr\"{o}dinger operator. For the upcoming discussion, it is useful to keep in mind the following concrete form of $K(x)$
\begin{equation}
    K(x) = \begin{cases}
        (1-x^2)^\theta &\text{ for } x \in (-1,1)\\
        0 &\text{ for } x \in(-\infty,-1] \cup [1,\infty)
    \end{cases} \label{eq:K_theta}
\end{equation}
\Cref{eq:K_theta} undergoes the spectral transition as described above. \cref{fig:K_theta} shows the behavior of \cref{eq:K_theta} at various representative values.

\subsubsection{Semiclassical intuition from the freezing time}
To understand the spectral transition of \cref{eq:2_reduced_simplified}, let us begin with a review of the classical trajectories of two fractons. 
A two-body trajectory is shown in \cref{fig:quantum_sketch}(a). 
We see that generically two closely separated particles (necessary for the energy to be non-zero) separate out to the edges of their `Machian reach' allowed by the support of the pair inertia function $K(x)$ and eventually freeze in position as the momenta diverge. 
The time taken for the particles to separate and become motionless in position, which we will refer to as the classical freezing time $\tau_f$, can be computed by integrating Hamilton's equations as~\cite{AP2023NRFractons} 
\begin{equation}
    \tau_f \propto \int_{x_0}^1\frac{dx}{\sqrt{ K(x)}}. \label{eq:tauf}
\end{equation}
For any $|x_0| < 1$, if $K$ vanishes as $(1-|x|^2)^{\theta}$ at the edges $x=\pm 1$, $\tau_f$ \emph{diverges} for $\theta > 2$  and \emph{converges} for $\theta < 2$.
Hence, classically, there is a transition in dynamics as $\theta$ crosses $2$ from bounded to unbounded motion in time. The latter resembles a free particle that typically hosts a continuous spectrum, while the former is a bound particle with a discrete spectrum, suggesting a spectral transition at $\theta = 2$. Of course, this quantum intuition comes from experience with the behavior of the Schr\"{o}dinger operator which our Hamiltonian in \cref{eq:2_reduced_simplified} is decidedly not. 

\subsubsection{Self-adjointness}
For a quantum Hamiltonian to represent a physical observable, it must be a self-adjoint operator on a Hilbert space---in our case, the space of square-integrable functions $L^2([-1,1])$ defined on the coordinate space $x \in [-1,1]$. The importance of this condition stems from the spectral theorem, which guarantees that a self-adjoint operator has a real spectrum and a complete set of eigenfunctions.

The formal differential expression $\hat{H} = -\frac{d}{dx} K(x) \frac{d}{dx}$ is not a complete operator until its domain of definition, $\mathcal{D}(\hat{H}) \subset L^2([-1,1])$, is specified. We begin by requiring $\hat{H}$ to be symmetric (or Hermitian), meaning $\braket{\hat{H} \psi}{\phi} = \braket{\psi}{\hat{H}\phi}$ for all $\psi, \phi \in \mathcal{D}(\hat{H})$. Integrating by parts yields a boundary term:
\begin{equation}
\left[-K(x) \overline{\psi'(x)}\phi(x) + K(x) \overline{\psi(x)}\phi'(x)\right]_{-1}^{1}.
\label{eqn:boundary_term}
\end{equation}
If we initially choose a minimal domain, for instance, of smooth functions that vanish at the boundaries $x=\pm 1$, this boundary term is zero, and $\hat{H}$ is symmetric.

However, a symmetric operator is not necessarily self-adjoint~\cite{reed_simon1}. An operator is self-adjoint if its domain $\mathcal{D}(\hat{H})$ is equal to the domain of its adjoint, $\mathcal{D}(\hat{H}^\dagger)$. A symmetric operator for which this is not true may have one, many, or no self-adjoint extensions, each corresponding to a different choice of boundary conditions that makes the physics well-defined. The number of such extensions is determined by the behavior of the solutions to the eigenvalue equation near the boundaries.

For the operator in \cref{eq:2_reduced_simplified}, this classification is provided by SL theory and depends critically on the exponent $\theta$:
\begin{itemize}
    \item For $\theta > 2$, the operator is in the \emph{limit-point} case at both boundaries $x=\pm 1$. This means the minimal symmetric operator is \emph{essentially self-adjoint}: it has a unique self-adjoint extension. No additional boundary conditions need to be imposed to define its domain. Physically, the dynamics are well-posed without specifying rules at the boundary. As we will see, this case leads to a continuous spectrum.
    
    \item For $\theta < 2$, the operator is in the \emph{limit-circle} case. The minimal operator is not essentially self-adjoint and admits an infinite family of self-adjoint extensions. To specify a physical Hamiltonian, one must choose a particular extension by imposing a boundary condition on the functions in its domain. This choice is what leads to a discrete spectrum.
\end{itemize}

\subsection{Discrete spectrum}
\label{sec:discrete_spectrum}
\subsubsection{Self-adjoint boundary conditions}
For the limit-circle case ($\theta < 2$), the deficiency indices of $\hat{H}$ are $(n_+, n_-) = (2,2)$, since for any non-real energy $E$, the equation $\hat{H}\psi = E\psi$ has two square-integrable solutions on $[-1, 1]$. This implies~\cite{reed_simon1} that the self-adjoint extensions are parameterized by a $2 \times 2$ unitary matrix, or four real parameters.

A known result (Proposition 10.4.2 of \cite{zettl2005sturm}) for SL operators of this form states that the self-adjoint extensions are parameterized by the following boundary conditions:
\begin{equation}
\label{eq:general_BC}
A \begin{pmatrix}
(-K \psi')(-1) \\
(\psi -vK\psi')(-1)
\end{pmatrix}
+ B
\begin{pmatrix}
(-K \psi')(+1) \\
(\psi - vK\psi')(+1)
\end{pmatrix}
=
\begin{pmatrix}
0 \\
0
\end{pmatrix},
\end{equation}
where $v(x) = \int_0^x\frac{1}{K(x')} dx'$, and $A$ and $B$ are $2\times 2$ matrices such that:
\begin{equation}
A \begin{pmatrix}
0 & -1 \\
1 & 0 \\
\end{pmatrix} A^{\dagger}
= B \begin{pmatrix}
0 & -1 \\
1 & 0 \\
\end{pmatrix} B^{\dagger}.
\end{equation}
Physically, we are typically interested in uncoupled boundary conditions that do not mix the behavior at $x=-1$ and $x=+1$. A simple and physically motivated choice is to set $A = \begin{pmatrix} 1 & 0 \\ 0 & 0 \end{pmatrix}$ and $B = \begin{pmatrix} 0 & 0 \\ 1 & 0 \end{pmatrix}$, which leads to the boundary condition:
\begin{equation}
(K \psi')(\pm 1) = 0.
\end{equation}
The domain restriction to functions satisfying this boundary condition defines one possible self-adjoint Hamiltonian. Another class of uncoupled boundary conditions can be parameterized by a single parameter $\beta$:
\begin{equation}
\label{eq:uncoupled_BC}
\left[ \cos(\beta)(-K \psi') + \sin(\beta)(\psi -vK\psi') \right](\pm 1) = 0.
\end{equation}

For all systems with $\theta < 2$, such as $K = (1 -x^2)^{\theta}$, the choice of such boundary conditions leads to a discrete spectrum. This has implications on the qualitative dynamics of a wavepacket, which will have no classical analog. To illustrate the dynamics, we now switch to focus on an exactly solvable case.

\subsubsection{Exactly solvable case}
For $\theta = 1$, where we expect a discrete spectrum, we can exactly solve the Schr\"{o}dinger equation.
We take $K = 1-x^2$, which vanishes linearly at $x = \pm 1$.
Writing $E_l = l(l+1)$, we have
\begin{equation}
(1-x^2)\psi'' - 2x \psi' + l(l+1)\psi = 0,
\label{eq:legendre equation}
\end{equation}
which is exactly the Legendre differential equation.
We denote the energies by $E_l$ with integer $l \geq 0$.
For each energy, there are two independent solutions, $P_l(x)$ and $Q_l(x)$, Legendre functions of the first and second kind of degree $l$.
The $P_l(x)$  are finite at $x=\pm 1$, whilst the $Q_l(x)$ diverge logarithmically.
With the boundary condition $[K(x)\psi'(x)](\pm1) = 0$, the $Q_l(x)$ solutions are ruled out, so the orthonormal eigenstates are $\psi_l(x) = \sqrt{\frac{2l+1}{2}} P_l(x)$.

We stress the $\theta=1$ case as it describes the essential features of all $\theta < 2$ systems.
The classical system with $\theta < 2$ has a pair of particles separating out to $x_1 - x_2 = \pm 1$, and \emph{staying there}.
Contrast this to the quantum case:
the system has a discrete spectrum, with well-behaved, non-diverging eigenstates.
Therefore, a wave packet initialized with the two particles close to each other will never fully separate out to $x = \pm 1$. Thus quantization introduces new physics for $\theta < 2$.

\subsection{Liouville transform}
\label{sec:liouville_transform}

Let us focus on the finite energy spectrum $\{E\}$ which is governed by the SL operator defined in a finite interval.
\begin{equation}
     -\frac{d}{dx} \left( K(x) \frac{d \psi(x)}{dx}   \right)  = E \psi(x), ~~x\in(-1,1). \label{eq:2particles_H_interval}
\end{equation}
We assume that $K(x)$ vanishes at the edges of the domain as $\abs{x-x_{\mathrm{edge}}}^\theta$. 
We now employ the so-called Liouville transformation~\cite{hille1969lectures} to recast \cref{eq:2particles_H_interval} into an ordinary QM particle problem.
To do this, the position $x$ is transformed into $\xi(x)$, and the field $\psi$ is rescaled to $\Psi(\xi)$
\begin{equation}
    \xi(x) = \int_{0}^x \frac{ds}{\sqrt{K(s)}},~~~ x \in (-1,1). \label{eq:xtoxi}
\end{equation}
\begin{align}
    \Psi(\xi) &=    \left( K(x(\xi))\right)^{\frac{1}{4}} \psi(x(\xi)).
\end{align}
We also introduce the effective potential
\begin{equation}
V(\xi) = \frac{\partial_{\xi}^2 \left[ K(x(\xi))^\frac{1}{4} \right]}{K(x(\xi))^{\frac{1}{4}}}.
\label{eq:effective_potential}
\end{equation}
Pulling this altogether, we arrive at the familiar Schr\"{o}dinger equation for an ordinary quantum particle:
\begin{equation}
    -\frac{d^2 \Psi(\xi)}{d\xi^2} + V(\xi) \Psi(\xi)  = E \Psi(\xi), ~~\xi \in (-a,a),\label{eq:LiouvilleNormalForm}
\end{equation}
with $a \equiv \xi(1)$.
Crucially, the Liouville transform is unitary and can be quickly verified by checking $\langle \phi_1, \phi_2\rangle_\xi = \langle \psi_1, \psi_2 \rangle_x$, so that dynamics evolved on $\Psi(\xi, t)$ under the new (ordinary QM) Hamiltonian corresponds to the exact same dynamics on $\psi(x,t)$.
The Liouville transform directly provides the tools to prove two important results:
\begin{enumerate}
    \item The existence of a spectral transition at $\theta = 2$.
    \item The pile up of probability at the edges for $\theta > 2$, and standard boundary reflection for $\theta < 2$.
\end{enumerate}

\subsubsection{Spectrum transition}
Marvelously, we have $a \propto \tau_f$ defined in \cref{eq:tauf}, and therefore as $\theta$ is tuned across $\theta = 2$, the domain in which the Schr\"{o}dinger operator is defined changes from a finite interval where we can expect a discrete spectrum, to an infinite interval where we can expect a continuous spectrum.
This, however, depends on the behavior of $V(\xi)$, especially at the ends of the domain. 

We focus on the two different cases, either side of $\theta = 2$.
For $\theta > 2$, the finite range $x \in (-1, 1)$ is mapped to $\xi \in (-\infty, \infty)$.
From \cref{eq:effective_potential}, $V(x(\xi)) \sim (1 - \abs{x})^{\theta-2}$ as $x \rightarrow \pm 1$, hence $V \rightarrow 0$ as $\xi \rightarrow \pm \infty$: the potential vanishes at the edges.
Following intuition for quantum mechanical problems defined on an infinite line, the Liouville transform then correctly predicts a continuous spectrum for energies in the $\theta > 2$ case.

The $\theta < 2$ case is more nuanced, as $\xi$ is mapped onto a finite range.
However, for some $\theta$, the potential $V(\xi)$ \emph{diverges} at the edges---does this imply a continuous spectrum or a discrete one? 
The key lies in the boundary condition defined on $\psi(x)$, and how it maps to a different condition for $\Psi(\xi)$.
We focus on $\theta = 1$ here for concreteness.
We now need to map the canonical boundary condition $[K(x)\frac{d}{dx}\psi(x)](\pm1) = 0$. 
Transforming between the old and new coordinates with $\xi = \arcsin(x)$ and $\Psi(\xi) = K(x)^{1/4} \psi(x)$, we have that
\begin{equation}
K \psi'(x) = 0 = -\frac{1}{4}K^{-\frac{1}{4}} K' \Psi + K^{\frac{1}{4}} \Psi'.
\end{equation}
Expanding about $\xi = \pi/2$ (i.e., $x = 1$), let $\xi = \pi/2 - \epsilon$, then $x \approx 1 - \epsilon^2/2$, so $K(x) = 1-x^2 \approx \epsilon^2$. The boundary condition becomes
\begin{equation}
\epsilon^{-\frac{1}{2}} \Psi + 2\epsilon^{\frac{1}{2}} \Psi' \approx 0.
\end{equation}
For this to hold as $\epsilon \rightarrow 0$, we simply require that the wavefunction vanishes at the boundary:
\begin{equation}
\Psi \left(\pm \frac{\pi}{2} \right) = 0.
\end{equation}

Solving the quantum mechanical problem for $\Psi(\xi)$ with this boundary condition, despite the diverging potential at $\pm \frac{\pi}{2}$, leads to a discrete spectrum, recovering the energy levels of the Legendre differential equation.
Importantly, this is true for all $\theta < 2$: a discrete spectrum is always obtained.

\subsubsection{Dynamics}

\begin{figure}[!h]
    \centering
    \includegraphics[width=8.6cm]{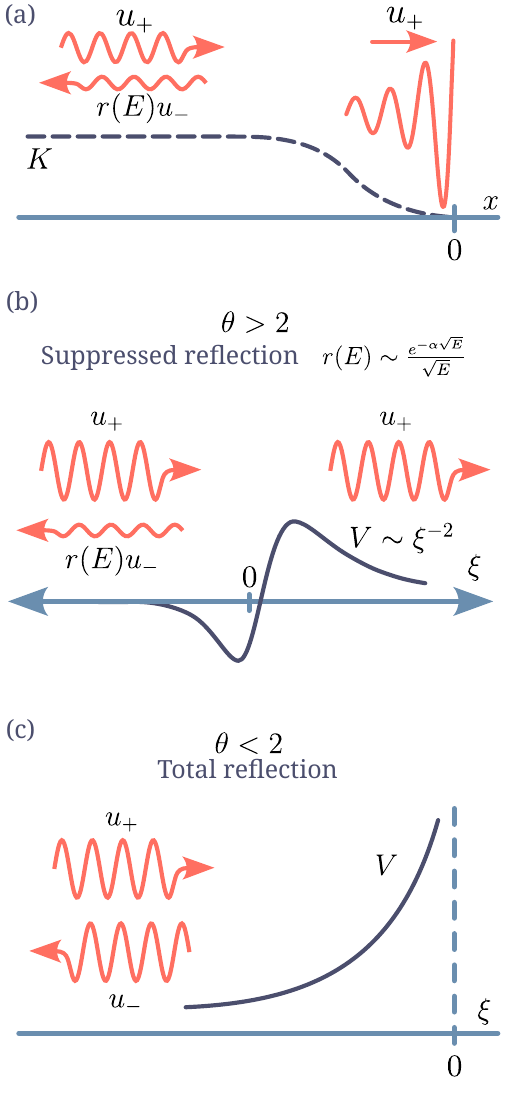}
    \caption{Wave reflection in the two regimes $\theta > 2$ (continuous spectrum) and $\theta < 2$ (discrete spectrum).
    (a): Incoming wave $u_+$ and reflection $r(E)u_-$, in the original fracton $x$-coordinates.
    (b): Liouville map onto $\xi \in (-\infty, \infty)$ for $\theta > 2$. 
    The effective potential $V$ decays and behaves nicely with no singularities. 
    An incoming wave with high enough energy is expected to pass through with suppressed reflection.
    (c): Liouville map for $\theta < 2$, with $x=0$ mapping onto a finite value of $\xi$.
    The effective potential becomes infinite at $\xi = 0$, so that any incoming wave will totally reflect.}
    \label{fig:reflection_sketch_liouville_map}
\end{figure}

The energy spectrum transitions from a discrete one to a continuous one as we cross $\theta =2$.
How does the dynamics change as this transition is crossed?
Once again we use the Liouville transform to tackle this problem, as the dynamics of an ordinary particle is easy to solve.

Comparing to the classical dynamics \cite{AP2024MachianFractons}, we might expect a wavepacket to pile up in amplitude at $x = +1$ or $-1$.
We wish to use the WKB states to study the quantum dynamics: how does a wavepacket evolve under $\hat{H}$, and does it match the classical behavior, or does the wavepacket reflect off the boundaries?
Consider for simplicity:
{\begin{equation}
K =        
\begin{cases}
\sim 1 & \text{for } x \lesssim -1\\
\sim \abs{x}^{\theta} & \text{near } x=0 \\
0 & \text{for } x \geq 0
\end{cases}
\end{equation}
A tractable function that realises this is 
\begin{equation}
K =        
\begin{cases}
\left(\frac{x^{\frac{\theta}{2}}}{1+x^{\frac{\theta}{2}}}\right)^2 & \text{for } x<0\\
0 & \text{for } x \geq 0
\end{cases}
\end{equation}
In \cref{fig:reflection_sketch_liouville_map}(b), we show the general form of $V(\xi)$ for $\theta > 2$, with $\xi \in (-\infty, \infty)$.
For all $\theta > 2$ (we treat $\theta < 2$ later), the potential $V$ tends to $0$ as $\sim \xi^{-2}$ when $\xi \rightarrow \infty$.
We have now reduced the fracton problem into a traditional particle-scattering problem!

The question we want to answer is as follows: consider an incoming wave in the original $x$ coordinates---does this reflect from the region of $K \rightarrow 0$, or does it pile up there, as in the classical case?
(\cref{fig:reflection_sketch_liouville_map}(a)).
Intuitively looking at the potential for $\theta > 2$ in \cref{fig:reflection_sketch_liouville_map}(b), the wave will mostly transmit through the barrier, with some small reflection which will depend on the wave energy $E$.
In simpler terms, we \emph{do} mostly expect waves to pile up at $\xi \rightarrow \infty$, or $x \rightarrow 0$.
We now seek to quantify this, using semiclassical analysis.

In standard WKB semiclassical analysis, where the length scale of momentum variations is much smaller than that of the length scale of the scattering potential $V$, we may expand a wave of the form $\psi \sim e^{iS}$ in powers of $\hbar$:
\begin{equation}
    \psi = \exp\left(\frac{1}{\hbar}S_0 + S_1 + \hbar S_2\right).
\end{equation}
Following this through, by matching orders of $\hbar$ in the Schr\"{o}dinger equation, leads to a ``position-dependent momentum'' of the form $p(x) = \pm \sqrt{E - V(x)}$.
This encapsulates two solutions, right- and left-traveling waves:
\begin{equation}
u_\pm(x) = \frac{1}{\sqrt{p(x)}} \exp \left\{\pm i \int^\xi p(\xi') d\xi' \right\} 
\end{equation}
While this captures the behavior of an incident wave slowing down, it is insufficient to model reflections, as reflections involve an essential singularity in $\hbar$, which is not going to be captured by a perturbative $\hbar$ expansion.
Nevertheless, standard techniques exist for lifting the WKB procedure, which we detail.

We commence by writing the Bremmer wavefunction~\cite{atkinson1960wave}, which decomposes the wavefunction into left- and right-ward traveling waves:
\begin{align}
\psi(\xi) &= \psi_+(\xi) + \psi_-(\xi) \nonumber\\
&= b_+(\xi) u_+(\xi) + b_-(\xi) u_-(\xi) 
\end{align}
We have now introduced the spatially varying $b_\pm(\xi)$.
As of yet, these functions are arbitrary, so $b_\pm(\xi)$ could contain any rapid oscillations.
Physically, we wish to impose sensible left- and right-wards traveling waves, and have the $b_\pm(\xi)$ pick up on reflection/transmission at all $\xi$.
To do so, we impose that the stationary probability current $j$ be split additively between a left piece and right piece: $j \sim j_\mathrm{right} - j_\mathrm{left} = \mathrm{const}$, with no cross-terms~\cite{heading2013introduction}.
Conventionally, this is equivalent to discretizing $V(\xi)$ into steps, and calculating transmission/reflection at each step \cite{berry1972semiclassical}, yet this may also be derived in the continuum \cite{atkinson1960wave}.
Either way, we arrive at two coupled first-order equations:
\begin{equation}
b_\pm'(\xi) = \frac{p'(\xi)}{2p(\xi)} \exp \left\{ \mp 2i \int^\xi p(\xi') d\xi' \right\}
\end{equation}
In principle, $b_\pm(\xi)$ are solved for iteratively, commencing with the initial values for an incoming right-wards wave from $\xi = -\infty$, and no left-wards wave from $\xi = + \infty$: $b_+(-\infty) = 1$ and $b_-(+\infty) = 0$.
Seeking to answer our primary question, we must determine the reflected component $r(E)$ at $\xi = -\infty$, i.e.\ what bounces back in the original $x$-coordinates from $x = 0$?
Integrating with the boundary conditions above, we find that
\begin{equation}
r(E) = b_-(-\infty) = \int_{-\infty}^{\infty} \frac{p'}{2p} e^{2i \int^\xi p(\xi') d\xi'}  d\xi.
\end{equation}
In the first-order approximation, we take $b_+(\xi) \approx 1$ everywhere.
Further, we assume a high energy $E \gg V_\mathrm{max}$ for the incoming wave, such that $p(x) = \sqrt{E - V(x)} \approx \sqrt{E}$.
Hence, in this first-order and high-energy approximation, we simply have:
\begin{equation}
r(E) = b_-(-\infty) = -\frac{1}{4E} \int_{-\infty}^\infty V'(\xi) e^{2 i \sqrt{E} \xi } d\xi,
\end{equation}
where we then use integration by parts and $V(\pm \infty) = 0$ to arrive at~\footnote{Generally speaking, we must iterate this procedure, each time picking up an additional integral: $\int d\xi_1 \int d\xi_2 \cdots \int d\xi_n$, and so on.
Physically, each integral corresponds to a point where the wave reflects.}:
\begin{equation}
r(E) = \frac{i}{2\sqrt{E}} \int_{-\infty}^\infty V(\xi) e^{2i \sqrt{E} \xi} d\xi.
\end{equation}
To understand the reflection behavior, we must determine the energy dependence of $r(E)$.
The integral is simply a Fourier transform, so $r(E) \sim \frac{1}{\sqrt{E}} \tilde{V}(E)$.
$V$ is a well-behaved real function, with no real singularities, and decays as $\sim \xi^{-2}$.
Hence, $\tilde{V}(\sqrt{E})$ decays exponentially in $\sqrt{E}$ with the distance to the nearest complex singularity:
\begin{equation}
r(E) \sim \frac{e^{-\alpha \sqrt{E}}} {\sqrt{E}} 
\end{equation}
We now conclude the mathematics and remark upon the physics: at moderate energies, the reflection is completely negligible.
Mapping back to our original fracton problem in $x$, we are led to a striking conclusion.
For $\theta > 2$, incoming waves into points of $K \rightarrow 0$ do \emph{not} reflect, even if there is a finite distance to $K = 0$.

We can take this further: consider a finite system, for example $K = (1 -x^2)^\theta$ on $x \in (-1, 1)$.
A generic wavepacket starting anywhere with some momentum kick will always pile up in probability at one of the edges---this is in direct correspondence to the unusual classical dynamics!

Throughout this section, we have assumed the $\xi$ is mapped onto an infinite form, and the effective potential $V$ looks like \cref{fig:reflection_sketch_liouville_map}(b).
This is true for $\theta > 2$ only.
We now turn to the $\theta < 2$ case, which would map the finite interval $x \in (-1, 1)$ to a finite interval in $\xi$.
The story is as shown in \cref{fig:reflection_sketch_liouville_map}(c): at the edge of $\xi$, the effective potential diverges.
Then, the conclusion is simple: an incoming wave will fully reflect.
Back in the original $x$-coordinates, we see familiar quantum behavior: waves incident on the $K \rightarrow 0$ point will reflect, instead of piling up.

\subsection{Reflection time}

From semiclassics, we have shown that wavepackets do not reflect off the Machian edges $K \rightarrow 0$ for $\theta > 2$.
To confirm this, we discretize the system (as in \cref{sec:lattice_models}) and study wavepacket dynamics.

To numerically verify the absence of reflections for $\theta > 2$, we consider a rightwards-traveling wavepacket initialized in the system, and evaluate the reflection time. As shown in \cref{fig:reflection_time}, 
the reflection time, being a discretization-size effect,  is finite for any finite lattice spacing $a$. 
However, it increases in value with smaller lattice spacing. 
Hence, we expect in the continuum limit $a \rightarrow 0$, the reflection time will be infinite, consistent with our previous semiclassical calculations.
Essentially, we observe $r(E) \approx 0$, as the wavepacket in \cref{fig:reflection_time} has high energy, exponentially suppressing $r(E)$.

\begin{figure}[!htbp]
\centerline{\includegraphics[width=8.6cm]{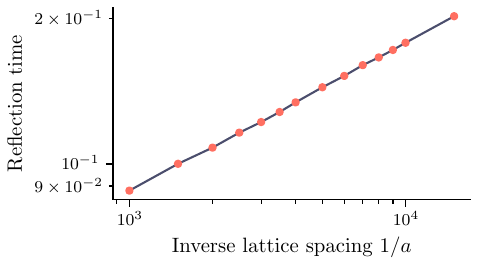}}
\caption[]{\label{fig:reflection_time} Numerical measurements for the wavepacket reflection time for $K = (1-x^2)^3$, i.e., $\theta=3$, where the Hamiltonian spectrum is continuous. The reflection time is defined as the time for $\langle \abs{x}^2 \rangle$ to peak. The initial wavepacket is taken to be $\psi(x, t=0) \sim \exp(-x^2/\sigma^2) \exp(i q x)$, with $\sigma=0.05$ and $q = 30$. In numerical simulations of wavepacket dynamics, the wavepacket piles up at one of the edges $\pm 1$, before reflecting. The time to reflect increases with the number of lattice discretization steps, implying that no reflections are present in the continuum limit. }
\end{figure}

\section{Three fractons}

\label{sec:three_fractons}
\subsection{The various blocks of the three-particle Hamiltonian}

Having fully characterized two fractons, we now consider the three-fracton system.
In reduced coordinates, the Hamiltonian is
\begin{multline}
    H = 
  -\left(\sqrt{2} \partial_{q_1}\right)K\left(\sqrt{2} q_{1}\right) \left(\sqrt{2} \partial_{q_1}\right) \\
  - \left(\frac{\partial_{q_{1}} + \sqrt{3} \partial_{q_2} }{\sqrt{2}}\right) K\left(\frac{{q_{1}} + \sqrt{3} {q_2} }{\sqrt{2}}\right) \left(\frac{\partial_{q_{1}} + \sqrt{3} \partial_{q_2} }{\sqrt{2}}\right) \\
 - \left(\frac{\partial_{q_{1}} - \sqrt{3} \partial_{q_2} }{\sqrt{2}}\right) K\left(\frac{{q_{1}} - \sqrt{3} {q_2} }{\sqrt{2}}\right) \left(\frac{\partial_{q_{1}} - \sqrt{3} \partial_{q_2} }{\sqrt{2}}\right) \label{eq:3_fracton_hamiltonian_maintext}
\end{multline}

We begin by pointing out that the dominant part of the three-particle Hilbert space consists of zero-modes, for any wavefunction with full support in the shaded region of \cref{fig:zero_mode}(b). 
Furthermore, the part of the Hamiltonian with non-zero energy, represented by the unshaded region in \cref{fig:zero_mode}(b), is block-diagonal. 
It was shown that under classical dynamics, for any generic initial conditions starting \emph{inside} the central hexagonal region, circumscribing the star of David in \cref{fig:zero_mode}(b), the trajectory eventually ends up at one of the six inward-facing triangles outside the star. 
Whereas finite-energy initial conditions \emph{outside} the hexagon, within the infinite strips emanating outward in \cref{fig:zero_mode}(b), correspond to a single fracton isolated from a close-separated pair, so the dynamics of the pair reduces to the two-particle problem. 
It is easy to verify that for the Hamiltonian in \cref{eq:3_fracton_hamiltonian_maintext}, the `genuinely three-particle' hexagonal region is block-diagonal with the central region, forming a distinct block from the rest. 
Furthermore, we have distinct blocks for each effective two-body problem, fixed by the separation of the pair from the singlet, represented by a line within the unshaded strips, perpendicular to the boundary.
It is clear that the finite-energy spectrum is dominated by these strips. 
Since the two-particle problem has already been solved, we only need to focus on the central hexagon to get the three-particle spectrum. 

In Ref~\cite{AP2023NRFractons}, it was shown that three classical fractons starting at close separation generically evolve to late-time attractor states, with two fractons oscillating about each other, and one particle freezing out. 
We now investigate to what extent these unusual dynamical features carry over into quantum fractons.

For the two-body system, we made headway by applying results of Sturm--Liouville (SL) theory and WKB semiclassical methods. 
For the three-body case, however, there is no well-established, off-the-shelf extension of SL theory to our setting. 
The Hamiltonian block represented by the central hexagon in \cref{fig:zero_mode}(b) is itself partitioned by \emph{internal} co-dimension-one loci (dashed lines) on which one of the three pair inertia functions $K$ vanish as a pair of particles go out of Machian reach. 
Across each such ``internal line'' the wavefunctions must be matched on either side: see \cref{sec:three_fracton_internal_lines}. 
Unlike the two-body endpoints $x=\pm1$, where standard SL theory prescribes natural self-adjoint boundary conditions, a general SL-like framework that classifies all admissible matching conditions for this multi-directional, degenerate problem is, to our knowledge, unavailable.

We therefore proceed via a lattice regularization that enforces dipole conservation. 
In the continuum limit, this construction canonically selects a particular set of matching conditions---equivalently, a particular continuum.
This choice lets us study genuine three-body physics non-perturbatively. In what follows we analyze the induced continuum operator numerically, focusing on low-lying eigenfunctions (which concentrate along the classical attractor lines) and on wavepacket dynamics, while \cref{sec:lattice_models} details the discretization and \cref{sec:three_fracton_internal_lines} discusses the associated matching conditions on the internal $K=0$ lines.

\begin{figure}[htbp]
\begin{overpic}[width=8.6cm,tics=10]{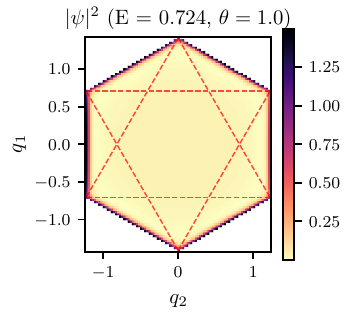}
     \put(49,48){\color[HTML]{FF6F61}\huge $\mathbf{3}$} 
     \put(49,68){\color[HTML]{FF6F61}\LARGE $\mathbf{2}$}
     \put(61.5,66){\color[HTML]{FF6F61}\Large $\mathbf{1}$}
\end{overpic}
\caption[]{Low-energy eigenfunction for $\theta = 1.0$, localized almost entirely within the $\mathbf{1}$ region of the hexagon: coinciding with where classical trajectories tend to at late times. 
This feature is typical of most low energy states. 
The Hamiltonian is discretized on a triangular lattice (aligning the three kinetic terms along principal axes), then we interpolate onto a square grid to produce the figure. 
Only coordinates within the hexagon are shown, as outside coordinates trivially correspond to two-fracton or one-fracton problems, and the Hamiltonian is block diagonal between the two regions.
The red dashed lines separate regions ($\mathbf{3}$, $\mathbf{2}_{a-f}$, $\mathbf{1}_{a-f}$), where region $\mathbf{3}$ has 3 of the $K$ functions being non-zero, and so on. 
}
\label{fig:numerical_3_fracton_wavefunction}
\end{figure}

\subsection{Low-lying wavefunctions}

First, we demonstrate general properties of three fractons by numerically solving for the spectrum, shown in \cref{fig:numerical_3_fracton_wavefunction}.
Strikingly, for generic $\theta$, low energy wavefunctions are almost entirely localized in the $\mathbf{1}$ regions, analogously to classical late-time states.
The lowest energy wavefunctions, for example the one in \cref{fig:numerical_3_fracton_wavefunction}, lie almost completely to the edge of the hexagon, corresponding to the steady-state of classical solutions where two particles separate out maximally from the third and settle down to indefinite oscillations.
As energies are increased, the wavefunctions are brought inwards, corresponding to smaller oscillations.

Notice the similarity to the two-fracton case: late-time classical states correspond to the quantum wavefunctions.
A crucial difference is that in the classical case, trajectories settle down to oscillate in one of the six $\mathbf{1}_{a-f}$ regions.
However, in the quantum version, wavefunctions overlap between the different $\mathbf{1}$ regions, implying \emph{tunneling} between the $\mathbf{1}_{a-f}$ regions---the quantum fractons tunnel between different permutations!

As is evident from \cref{fig:zero_mode}(b), the three-body Hamiltonian  \cref{eq:3_fracton_hamiltonian_maintext} has the symmetries of a hexagon. This is generated by the permutations of the three identical particles and spatial reflections, which appears in reduced coordinates as symmetries of the hexagon. The result of tunneling is that the eigenstates would organize according to the irreducible representations (irreps) of this symmetry group. For the hexagon, this is a so-called Dihedral group which has only one and two dimensional irreps. Thus, we expect any eigenstate to be either a singlet or doublet with different charges. For more details on symmetries see \cref{app:hexagon_symmetries}.

\subsection{Spectral transition}
We have shown rigorously that, for two fractons, there is a discrete-to-continuum spectral transition as $\theta$ crosses $\theta=2$.
This transition is accompanied by a dramatic change in the dynamics, with $\theta>2$ leading to wavepackets piling up at $\pm1$ in an apparent mirroring of classical fracton behavior.
We now seek to numerically study the three-fracton behavior and find evidence for a spectral transition.

\begin{figure}[htbp]
\centerline{\includegraphics[width=8.6cm]{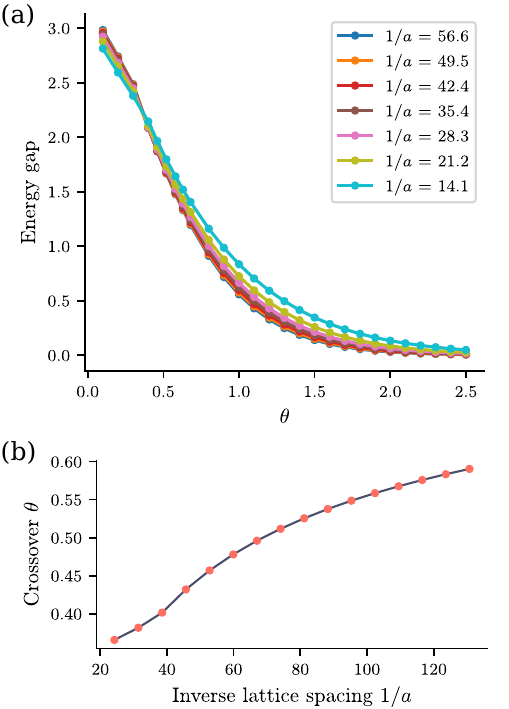}}
\caption[]{(a): $E_1 - E_0$ energy gap spacing for varying $\theta$. 
A crossover is found at some $\theta_c$, implying a continuous energy spectrum for $\theta > \theta_c$ as the discretization goes to zero.
Some excess data points are excluded in (a) for clarity, as the plots for small discretizations look identical, but all of our data has been processed in (b).
(b): Scaling for the crossovers in (a). 
The crossing value of $\theta_c$ drifts to higher values at smaller discretizations.
We conjecture in the main text that $\theta_c$ flows to $\theta_c = 2$ in the continuum limit.
}
\label{fig:3_fracton_crossover}
\end{figure}

We work in the triangular lattice discretization of the reduced $q_1$--$q_2$ coordinates and take $K=(1-x^2)^\theta$ as usual.
To test for a transition in $\theta$, we compute the energy spacing between the first two energy eigenvalues and perform a scaling as a function of the discretization lattice constant $a$.
We show the gap versus $\theta$ in \cref{fig:3_fracton_crossover}(a) for various $a$. We also show the drift of the critical value  $\theta_c$ separating continuous from discrete spectrum, estimated from crossing of curves for closest values of $a$, in \cref{fig:3_fracton_crossover}(b).

For large values of $a$, a crossing in energy levels occurs for $\theta_c \sim 0.4$, implying a continuous energy spectrum for $\theta > \theta_c$.
Finite size scaling of the crossing indicates that $\theta_c$ \textit{increases} with the level of discretization.

We conjecture that in the continuum limit, the three-fracton transition imitates the two-fracton transition, such that $\theta_c \rightarrow 2$.
We leave a complete numerical study of $\theta_c$ as a task for future work.
Assuming this conjecture to be true, the close-separated three-fracton system is characterized by the ``microscopic'' two-fracton transition at $\theta=2$, itself originating from the $\theta=2$ timescale divergence for classical fractons. 
However, let us remind ourselves that, as discussed earlier, the finite-energy part of the three-particle spectrum is dominated by the effective two-particle sectors i.e.\ the infinite strips of \cref{fig:zero_mode}(b). 
So we can be certain that the system \emph{as a whole} undergoes a spectral transition at $\theta_c = 2$ irrespective of the behavior of the subdominant central hexagonal block.

\subsection{Dynamics}

\begin{figure*}[!htbp]
    \centering
    \includegraphics[width=\textwidth]{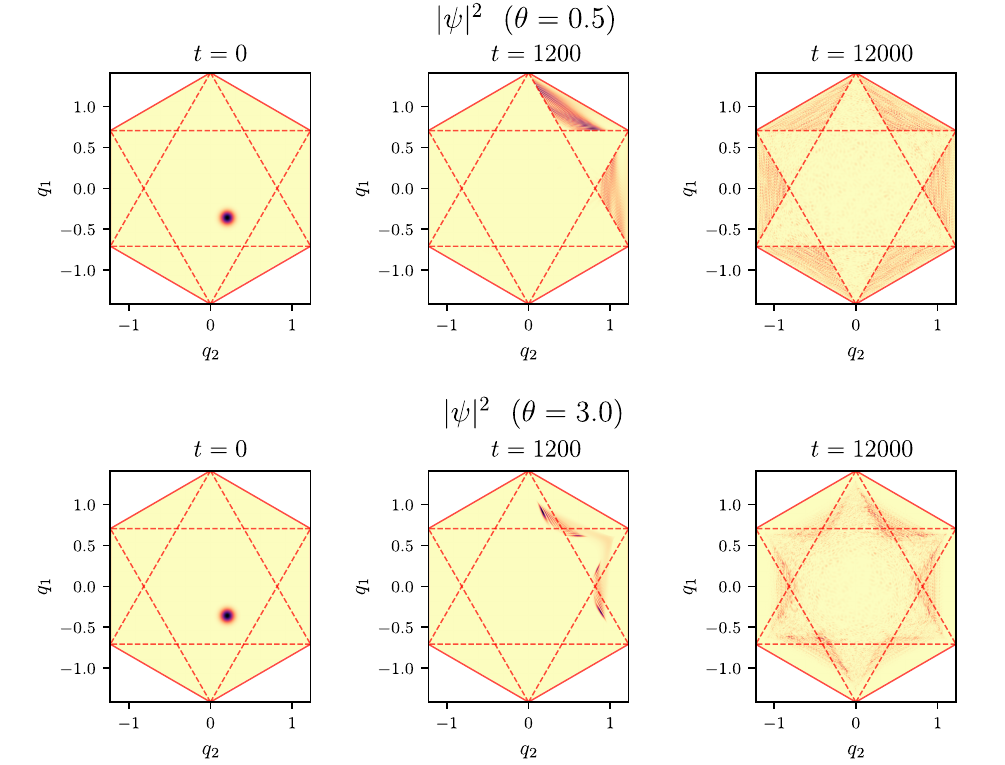}
    \caption{
    Three-fracton wavepacket evolution. 
    The starting wavepacket is taken to be some localized wavepacket in the $\mathbf{3}$ region with a momentum kick, the exact form being $\psi(x, t=0) = \exp(- (x_1^2 + (x_2-0.5)^2 + x_3^2) / (0.1)^2) \cdot \exp(100i~(x_1 + 0.25x_2))$.
    This state is taken to represent a generic, non-fine-tuned wavepacket.
    The $K$-function is taken to be the usual $K(x) = (1-x^2)^\theta$. 
    \emph{Top:} Evolution for $\theta=0.5$, which is below the transition $\theta_c$, so the spectrum is discrete. 
    The wavepacket reaches the $\mathbf{2}$--$\mathbf{1}$ boundary, and passes through, ending up stuck in the $\mathbf{1}$ region. 
    This is analogous to three classical fractons!     \emph{Bottom}: Evolution for $\theta = 3.0$, conjectured to be above $\theta_c$, where the energy spectrum is a continuum.
    The wavefunction only partly enters the $\mathbf{1}$ during the run of the simulation.
    }
    \label{fig:3_fracton_evolution}
\end{figure*}

For \emph{classical} trajectories of closely spaced three fractons, there is \emph{no} distinction between $\theta < 2$ and $\theta > 2$: trajectories always end in the $\mathbf{1}$ region, after a finite time. Essentially, the timescale divergence at $\theta = 2$ is uniquely present in the two-fracton system.

In \cref{fig:3_fracton_evolution}, we show Hamiltonian evolution for wavepackets for $\theta = 0.5 < \theta_c$ and $\theta = 3.0$ (conjectured above $\theta_c$).
For $\theta=0.5$, the wavepacket shoots off rapidly into the $\mathbf{1}$ regions.
The wavefunction enters a superposition between different permutations, due to tunneling.
Importantly, this behavior mirrors the classical behavior, where oscillations in $\mathbf{1}$ are observed.

Now, it is interesting to contrast with the $\theta = 3.0$ dynamics (\cref{fig:3_fracton_evolution}).
We find that the wavefunction partly permeates into the $\mathbf{1}$ region, whilst mainly having support in the $\mathbf{2}$ regions.
The classical version has no distinction on either side of $\theta = 2$, so we believe the simulations are limited by computer time---given more time, we expect the wavepacket to primarily be localized in the $\mathbf{1}$ regions.
Nevertheless, the $\theta = 0.5$ and $\theta = 3.0$ dynamics are clearly distinct, with the $\theta = 3.0$ dynamics progressing much slower, likely stemming from the continuous spectrum for $\theta > \theta_c$.

\subsection{Order-by-disorder}

In our previous study of classical many-body fractons~\cite{AP2024MachianFractons}, we drew parallels to the phenomenon of classical order-by-disorder (OBD)~\cite{MoessnerChalkerOBDPhysRevLett.80.2929,ChalkerOBD2011} in the formation of clusters. We argued that fractons order into clusters, as these configurations maximize the number of flat directions of the energy by ``switching off'' as many of the $K(x_i-x_j)$ as possible. 
We now argue that our three-particle results above can be rationalized by a version of quantum order by disorder wherein the quantum fluctuations about a given point in a naive ground state manifold select a subset exhibiting the softest orthogonal directions~\cite{shender1982antiferromagnetic,henley1987ordering}. 

Specifically, let us consider wavefunctions restricted to the regions labeled $\mathbf{3}$/$\mathbf{2}$/$\mathbf{1}$, representing the number of $K(x_i-x_j)$ switched on in each region, as in \cref{fig:numerical_3_fracton_wavefunction}.
Intuitively, the OBD argument suggests that we should find the lowest energy states in Region $\mathbf{1}$ where the Hamiltonian has the maximum number of flat directions, in this case exactly one direction. 
This is indeed the case, as we showed above.

Making this precise is more delicate, as the different regions have different areas, and one needs to associate boundary conditions with the individual regions to talk sensibly about states localized to them. 
As a first pass, we solve the Hamiltonian numerically separately in each region, having switched off the terms that connect regions.
In the $\mathbf{1}$ region, we find a large number of zero-energy states, which can be chosen along infinitesimally thin lines, similar to in \cref{fig:numerical_3_fracton_wavefunction}. 
By contrast, in regions $\mathbf{2}$ and $\mathbf{3}$, we find a single zero-energy state and all other low-lying states at non-zero energies, with the lowest energy $\mathbf{3}$ state being higher than in $\mathbf{2}$. 
Now imagine switching back on the couplings between the different regions.
The low-lying spectrum then reveals itself: a single state at zero energy with a constant wavefunction and the next set of wavefunctions is dominated by the stripes in the $\mathbf{1}$ region, connected by tunneling through the $\mathbf{2}$ region.
This intuition generalized to an arbitrary number of fractons suggests that the possibility of clustering and ergodicity breaking exists even in the quantum regime.

\section{Comments on the many-fracton problem}
\label{sec:comment_on_many_fracton}

\subsection{Connection to lattice fractons}
We have argued that for three fractons, the Hamiltonian is block-diagonal. In particular, each effective two-particle sector, corresponding to dynamics of two particles with a \emph{fixed} separation from an isolated particle, represents a distinct block. These blocks arise from the locality of the pair inertia function $K$ even after the symmetries have been resolved. As such, they represent distinct \emph{Krylov} subspaces. For any finite number of fractons, $N$, no matter how large, an extension of the 3-fracton picture can be used to argue that such a block-diagonal structure persists. This appearance of a large number of Krylov sectors is reminiscent of `Hilbert Space Fragmentation' seen in lattice quantum fractons~\cite{KhemaniHermeleNandkishore_Shattering_PhysRevB.101.174204,SalaRakovskyVerresenKnapPollmann_FragmentationPhysRevX.10.011047}. Unlike the lattice however, the continuum Hilbert space is unbounded from above even for a finite number of particles making it more complex to use tools such as those in Refs~\cite{MorningstayKhemaniHust_Thermalization_PhysRevB.101.214205,SkinnerPozderac_Thermalization_2023,classenhowes2025universalfreezingtransitionsdipoleconserving}. The spectral transition found as we change the nature of the pair inertia function at the edges is not expected to be seen in lattice fractons. It is unclear how these features are modified in the thermodynamic limit at various finite densities of fractons. 

\subsection{Connection to field theory formulations}
In many-body physics, we often map systems into a continuum. Several recent studies have investigated continuum formulations of fractons --- field theories, often relativistic, with dipole conservation symmetry~\cite{Pretko_FractonGauge_PhysRevB.98.115134,GorantlaLamSeiberg_UVIR_PhysRevB.104.235116}. Does our system of fractons map onto such a (non-relativistic) field theory? Let us consider a Hamiltonian in $d$ spatial dimensions and write down the leading two terms in powers of momenta
\begin{multline}
	H = \lambda_1 \sum_{j , k=1}^N K_1(|\vec{x}_j-\vec{x}_k|)  \left(\vec{p}_j - \vec{p}_k \right)^2 \\+ \lambda_2\sum_{j , k=1}^N K_2(|\vec{x}_j-\vec{x}_k|) \left(\vec{p}_j - \vec{p}_k \right)^4  \label{eq:H_manybody}.
\end{multline}
Canonical quantization of \cref{eq:H_manybody} once again requires us to pick an operator ordering. For a particular choice, we can use the standard recipe~\cite{fetter2012quantum} to second-quantize the Hamiltonian to get a continuum bilocal field theory $H \rightarrow  \int d^dx d^d y \mathcal{H}(x,y)$. At this stage, the standard procedure is to expand the kernel in terms of distributions of increasing locality, leading with $K(x) \approx \delta(x)$ representing the ultra local limit. 
In field theories, renormalization group arguments often justify only keeping a delta function interaction, such as for $\phi^4$ interactions in a scalar field theory.
We emphasize that the results of this paper are at odds with this thinking --- it is dangerous to take the ultra-local limit for $K$ in fractonic systems.
If we naively take the ultra-local limit $K_{1,2}(x) \rightarrow \delta(x),$
\begin{equation}
    K_{1,2}(x) \rightarrow \delta(x), \label{eq:ultra_local}
\end{equation}
we get a local Hamiltonian density 
\begin{equation}
     \int d^dx d^d y ~\mathcal{H}(x,y) \xrightarrow{K_{1,2}(x) = \delta(x)} \int d^dx ~\mathcal{H}(x),
\end{equation}
where,
\begin{multline}
       \mathcal{H} = -\lambda_1 (\phi^\dagger)^2 \left(\phi\vec{\nabla}^2 \phi -  \vec{\nabla} \phi. \vec{\nabla} \phi\right)+ h.c.\\
    +4 \lambda_2 \left|\phi \partial_\alpha \partial_\beta \phi - \partial_\alpha \phi \partial_\beta \phi\right|^2. \label{eq:Pretko}
\end{multline}
$\phi$ is a complex scalar field and we have chosen to second-quantize the identical particles as bosons. We have also used Einstein's summation convention, so that repeated indices $\alpha, \beta = 1,\ldots,d$ are summed over. 
There are subtleties involving operator ordering in arriving at \cref{eq:Pretko}. Details can be found in \cref{app:continuum}.

\cref{eq:Pretko} is the non-relativistic Hamiltonian limit of the field theory derived by Pretko~\cite{Pretko_FractonGauge_PhysRevB.98.115134} from imposing dipole conservation on a field theory. Non-relativistic versions of other continuum formulations~\cite{GorantlaLamSeiberg_UVIR_PhysRevB.104.235116,Gromov_Multipole_PhysRevX.9.031035} can be derived from \cref{eq:Pretko}, for instance, by ignoring the amplitude fluctuations of the complex scalar $\phi \sim A e^{i \theta}$ and focusing only on the phase $\theta$. The important step in deriving the field theories was taking the ultra-local limit of \cref{eq:ultra_local}. For example, by parametrizing with the normalized $K$:
\begin{equation}
K_n^\theta(x;a) = \frac{(a^2 - x^2)^\theta}{\int_{-a}^a (a^2 - y^2)^\theta dy} \text{  for } -a\le x\le a,
\end{equation}
whose limiting form becomes the delta function
\begin{equation}
\lim_{a \rightarrow0} K_n^\theta(x;a)  = \delta(x).
\end{equation}
In taking the limit, one makes the implicit assumption that the only important aspect of the pair inertia function $K(x)$ is its behavior within the interval of its support. This is clearly dangerous. As we have discussed, the nature of the multi-particle spectrum depends sensitively on the edge behavior of $K(x)$ - namely on the parameter $\theta$.
The nature of low-energy modes change with the spectral transition, and should therefore modify the low-energy many-body physics, highlighting the dangers of this limit.
Said differently, the field theories of Refs~\cite{Pretko_FractonGauge_PhysRevB.98.115134,GorantlaLamSeiberg_UVIR_PhysRevB.104.235116,Gromov_Multipole_PhysRevX.9.031035} are effectively applicable to isolated fractons and miss the sensitivity to $K(x)$, which directs clustering behavior.

\section{In closing}

The study of classical, non-relativistic, continuum fractons~\cite{AP2023NRFractons,AP2024MachianFractons,babbar2025classicalfractonslocalchaos,sadki2025phasespacefractons} has revealed multiple unusual dynamical features, including position-velocity space attractors, and a generic 
breaking of ergodicity in the many-body regime. 
Building on this, we set out to study their quantization with a view to understand how much of the classical structure survives into the quantum problem.
For the two-body problem, we identify a spectral transition at $\theta = 2$, controlled by the edge behavior of the pairwise inertia function $K(x)\sim \lvert x - x_{\mathrm{edge}} \rvert^{\theta}$, separating a discrete ($\theta<2$) and continuous ($\theta>2$) spectrum. 
Dynamically, for $\theta < 2$, we find familiar wavepacket dynamics reflecting off boundaries, while for $\theta > 2$ the behavior changes dramatically, with the wavefunction piling up at the edges of $K(x)$. 
The location of the transition mirrors a classical transition in the time taken for the fractons to reach the edges.
For three particles, we considered a symmetry-preserving lattice discretization, providing numerical evidence that a spectral transition persists, likely with the critical value $\theta_c\simeq 2$ in the continuum limit. We further showed that wave packet dynamics mimics classical dynamics in localizing the probability density primarily near classical attractors at late times and that the low energy eigenstates also are dominated by attractor regions.

A natural next step in the three-body study would be to place it on rigorous mathematical footing, analogous to Sturm-Liouville theory for the two-body case. In the bigger picture, the many-body regime at finite density remains to be explored, with the central question of whether or not quantum effects restore ergodicity.
The order-by-disorder argument presented in this paper suggests ergodicity breaking at low temperatures that maximizes clustering, mirroring the classical case. By extending the comparison to classical fractons, we expect quantum ergodicity breaking at finite temperature---but showing this requires a proper computational framework to be created.
Further, the applicability of fracton hydrodynamics~\cite{Gromovetal_FractonHydro_PhysRevResearch.2.033124} to our system is downstream of the question of ergodicity and thermalization. 

Viewed in the broader fracton context, our systems are---as explained in \cref{sec:lattice_models}---continuum limits of lattice models with hopping of diverging range in lattice units. From this perspective, we see that while states with fully isolated fractons are robust to the ultraviolet details contained in the pair inertia function $K$, the remaining states care sensitively about such details. This is why working directly in the naive ultra-local continuum limit~\cite{GorantlaLamSeiberg_UVIR_PhysRevB.104.235116} is not informative about the fracton dynamics that we have discussed in this paper, with the pair-inertia function determining clustering properties and a spectral transition.

Overall, Machian fractons continue to surprise, with their unusual behaviors now extending into the quantum regime.

\section*{Acknowledgements}
We would like to thank Tuhin Ghosh for discussions on the Sturm-Liouville theory of operators, Jon Keating for discussions on semiclassical mechanics, Aryaman Babbar for discussions on fractons and for previous collaboration~\cite{babbar2025classicalfractonslocalchaos}, and Nathan Seiberg for introducing us to his work on fracton field theories \cite{GorantlaLamSeiberg_UVIR_PhysRevB.104.235116}.
The work of A.P. was supported by the European Research Council under the European Union Horizon 2020 Research and Innovation Programme, Grant Agreement No. 804213-TMCS  and the Engineering and Physical Sciences Research Council, Grant number EP/S020527/1. The work of S.L.S. was supported by a Leverhulme Trust International Professorship, Grant Number LIP-202-014.

\appendix

\section{Lattice Discretization and Boundary Conditions}
\label{sec:lattice_models}
The continuum fracton models studied in this work can be understood as the low-energy limit of discrete lattice models. Here, we outline this connection and clarify the role of the lattice formulation in determining the continuum boundary conditions. For the two-body system, the effective Hamiltonian in reduced coordinates is given by \cref{eq:2_reduced_simplified}: $\hat{H} = -\frac{d}{dx} K(x) \frac{d}{dx}$, where $x$ represents the relative position of the two particles. Throughout this appendix, the symbol $x$ and its variants (e.g., $x_i$, $x_1$, $x_2$) will refer to discretized lattice positions.

A lattice version of this operator can be constructed using a finite-difference scheme on the reduced coordinate $x$. On a lattice with sites indexed by $i$ and spacing $a$, a suitable discretization for the bulk of the system is:
\begin{equation}
\label{eq:2_body_bulk_eqn}
\begin{split}
H\psi_{i} =\frac{1}{4a^{2}}  [ -(K_{i-1} + K_i) \psi_{i-1} +\\ (K_{i-1} + 2K_i + K_{i+1}) \psi_i \\- (K_i + K_{i+1}) \psi_{i+1} ],
\end{split}
\end{equation}
where $K_i \equiv K(x_i)$. In the limit $a \to 0$, this difference equation correctly reproduces the continuum differential operator. The specification of the model is completed by defining the Hamiltonian's action on the boundary sites, say $i=0$ and $i=N$. The behavior at these edges is particularly important as it determines the boundary conditions of the continuum theory, which are essential for ensuring the self-adjointness of the Hamiltonian.

In our numerical work, we adopt a natural choice where the bulk formula \cref{eq:2_body_bulk_eqn} is also used at the edges, with the understanding that the function $K_i$ is zero for sites outside the defined interval (e.g., $i < 0$ or $i > N$). For the edge sites $i=0$ and $i=N$, this yields:
\begin{align}
H\psi_{N} &=\frac{1}{4a^{2}}  [ -(K_{N-1}) \psi_{N-1} + (K_{N-1}) \psi_N ], \nonumber\\
H\psi_{0} &=\frac{1}{4a^{2}}  [ -(K_{1}) \psi_{1} + (K_{1}) \psi_0 ].
\end{align}
This particular choice of boundary discretization is not arbitrary; in the continuum limit, it corresponds to imposing a specific boundary condition. For this formulation, the emergent condition is $(K \psi')(x)=0$ at each boundary.

It is important to recognize that this is just one of many possible lattice regularizations. Different physical scenarios can be modeled by altering the lattice Hamiltonian at the boundaries. One can realize other self-adjoint extensions of the continuum operator by introducing explicit boundary terms. For example, adding a large on-site potential $V_b |\psi_b|^2$ at a boundary site $b$ would energetically penalize any non-zero wavefunction there. In the limit $V_b \to \infty$, this enforces a Dirichlet boundary condition, $\psi(b)=0$. Similarly, other local terms involving gradients at the boundary could be constructed to yield Neumann or Robin boundary conditions. Thus, the choice of lattice boundary terms provides a systematic way to specify the physical boundary conditions of the resulting continuum theory.
  
The discussion so far has focused on the reduced coordinates for the two-body problem. To better understand the microscopic origin of this model, it is instructive to switch back to the general coordinates, i.e., the individual positions of the two particles, which we denote on the lattice by $x_1$ and $x_2$. Consider the following form for the Hamiltonian, acting on a basis state $\ket{x_1}\ket{x_2}$:
\begin{multline}
  H\ket{x_{1}}\ket{x_{2}} = \\
  ( K(x_{1}-1,x_{2}+1) + 2K(x_{1}, x_{2}) + K(x_{1}+1,x_{2}-1) ) \ket{x_{1}} \ket{x_{2}} \\
  - ( K(x_{1}-1, x_{2}+1)  + K(x_{1}, x_{2}) ) \ket{x_{1} - 1}\ket{x_{2} + 1}\\
  - ( K(x_{1}+1, x_{2}-1)  + K(x_{1}, x_{2}) ) \ket{x_{1} + 1}\ket{x_{2} - 1}, 
\end{multline}
\noindent
where we are ignoring overall factors.
First, consider the case where $K$ is maximally local: 
\begin{equation}
  K(x_{1}, x_{2}) = \delta_{x_{1}, x_{2}},
\end{equation}
leading to the Hamiltonian: 
\begin{align}
  H\ket{x_{1}}\ket{x_{2}} =
  2\delta_{x_1,x_2}\ket{x_1}\ket{x_2} \nonumber \\
  - \underbrace{\delta_{x_{1},x_{2}}(\ket{x_{1}-1}\ket{x_{2}+1} + \ket{x_{1}+1}\ket{x_{2}-1})}_{\textrm{Hopping outwards: Fig \ref{fig:hopping_K0}(a)}}\nonumber  \\
  - \delta_{x_{1},x_{2}-2}\ket{x_{1}+1}\ket{x_{2}-1} \nonumber \\
  - \underbrace{\delta_{x_{1},x_{2}+2}\ket{x_{1}-1}\ket{x_{2}+1}}_{\textrm{Hopping inwards: Fig \ref{fig:hopping_K0}(b)}} 
\label{eq:K_0_hopping_hamiltonian}
\end{align}
These fractons do not get very far.
The fractons may only hop outwards if two fractons are directly on top of each other.

\begin{figure}
\centering
\includegraphics[width=8.6cm]{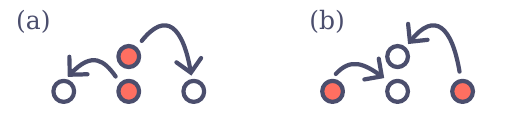}
\caption{Hopping terms generated by $K(x_{1}, x_{2}) = \delta_{x_{1}, x_{2}}$. The filled circles represent fractons occupying the lattice site at that position, the empty circles represent an unoccupied lattice site. The particles are distinguishable, so the Hamiltonian has four terms, from permutation of $x_{1} \leftrightarrow x_{2}$. These terms are shown in \cref{eq:K_0_hopping_hamiltonian}.}
\label{fig:hopping_K0}
\end{figure}

Increasing the range of K by one lattice site:
\begin{equation}
  K(x_{1}, x_{2}) = \delta_{x_{1},x_{2}} + \delta_{x_{1}, x_{2}+1} + \delta_{x_{1},x_{2}-1}
\end{equation}

Yields the following hopping Hamiltonian:
\begin{align}
  H\ket{x_{1}}\ket{x_{2}} = (\textrm{previous terms}) + \nonumber \\
  -\delta_{x_{1},x_{2}-1}\ket{x_{1}-1}\ket{x_{2}+1} \nonumber \\
  -\underbrace{\delta_{x_{1},x_{2}+1}\ket{x_{1}+1}\ket{x_{2}-1}}_{\textrm{Hop: Fig \ref{fig:hopping_K1}(a)}} \nonumber \\
  -\delta_{x_{1},x_{2}-3}\ket{x_{1}+1}\ket{x_{2}-1} \nonumber \\
  -\underbrace{\delta_{x_{1},x_{2}+3}\ket{x_{1}-1}\ket{x_{2}+1}}_{\textrm{Hop: Fig \ref{fig:hopping_K1}(b)}} \nonumber \\
  -2\delta_{x_{1},x_{2}-1}\ket{x_{1}+1}\ket{x_{2}-1} \nonumber \\
  -\underbrace{2\delta_{x_{1},x_{2}+1}\ket{x_{1}-1}\ket{x_{2}+1}}_{\textrm{Swap: Fig \ref{fig:hopping_K1}(c)}} \nonumber \\
   + \underbrace{(\delta_{x_{1},x_{2}+3} + 3\delta_{x_{1},x_{2}+1}+3\delta_{x_{1},x_{2}-1}+\delta_{x_{1},x_{2}-3})\ket{x_{1}}\ket{x_{2}}}_{\textrm{Diagonal term}} 
\label{eq:K_1_hopping_hamiltonian}
\end{align}

\begin{figure}
\centering
\includegraphics[width=8.6cm]{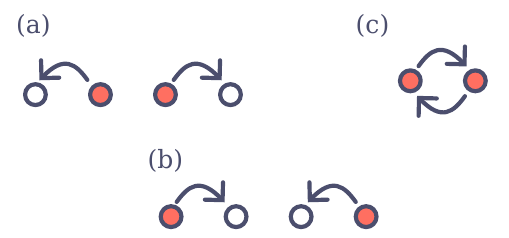}
\caption{Hopping terms generated by $K(x_{1}, x_{2}) = \delta_{x_{1}, x_{2}+1} + \delta_{x_{1},x_{2}-1}$. The corresponding terms in the Hamiltonian are listed in \cref{eq:K_1_hopping_hamiltonian}.}
\label{fig:hopping_K1}
\end{figure}

In general, including terms in $K$ of the form $\delta_{x_{1}, x_2+n} + \delta_{x_{1}, x_{2}-n}$ leads to the hopping terms in Fig \ref{fig:hopping_Kn}.
Generally, including terms in $K$ of the form $K \sim \sum\limits_{i=1}^{n}\delta_{\abs{x_{1}-x_{2}}, i}$ will allow mixing between states with the two particles up to a distance of $n$ lattice sites.
Viewing the lattice as a discretization, a $K(x)$ with support up to $\abs{x} \leq X$ will imply mixing between states up to $\frac{X}{a}$ sites, where $a$ is the lattice spacing.
We often require $K(x)$ to decrease with $x$, as imposed by locality, which would then decrease the hopping couplings at larger distances.

\begin{figure}[htbp]
\includegraphics[width=8.6cm]{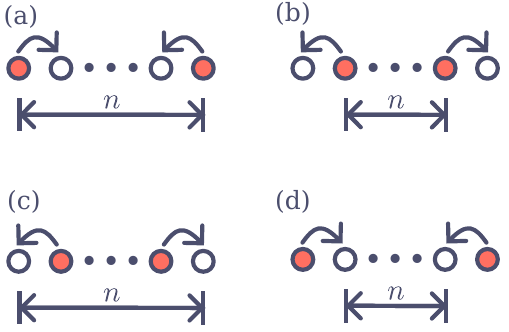}
\caption{Hopping terms generated by $K(x_{1}, x_{2}) = \delta_{x_{1}, x_{2}+n} + \delta_{x_{1},x_{2}-n}$. }
\label{fig:hopping_Kn}
\end{figure}

\section{Three fracton matching conditions}
\label{sec:three_fracton_internal_lines}
Consider again the three-body Hamiltonian in \cref{eq:3_fracton_hamiltonian_maintext}.
The behavior of the wavefunction across the red internal lines in \cref{fig:zero_mode}(b) must be considered carefully, as these lines correspond to when one of the three $K$ functions in the Hamiltonian goes to zero.
The points $x = \pm 1$ in the two-body case have $K \rightarrow 0$ and correspondingly a special boundary condition $K \psi' = 0$.
To derive an analogous matching condition, we consider $H\psi = E\psi$ close to a boundary line, for example a $\mathbf{2}$--$\mathbf{1}$ boundary:
\begin{align}
\label{eq:hamiltonian_near_2-1_boundary}
\begin{split}
H\psi = E\psi = 
  - \left(\sqrt{2} \partial_{q_1}\right)K\left(\sqrt{2} q_{1}\right) \left(\sqrt{2} \partial_{q_1}\right)\psi \\
  - \partial_\xi K\left(\frac{1}{\sqrt{2}}q_{1} + \sqrt{\frac{3}{2}} q_2\right) \partial_\xi \psi,
  \\
\partial_\xi =  \left(\frac{1}{\sqrt{2}}\partial_{q_{1}} + \sqrt{\frac{3}{2}} \partial_{q_2}\right).
\end{split}
\end{align}
There are two kinetic terms, with one going to zero: the $K\left( \sqrt{2} q_1 \right)$ term. 
The other can be approximated as constant, as it is continuous and non-zero.
Dropping constant factors in all the terms, we have 
\begin{equation}
    E\psi = -\partial_x (K(x) \partial_x \psi) - K_2 \left( \partial_x^2+ \partial_y^2 \right) \psi,
\end{equation}
where $x \equiv q_1 - Q_1$ ($Q_1$ being a constant), $y \equiv q_2$ and $K_2 \equiv K\left(\frac{1}{\sqrt{2}}q_{1} + \sqrt{\frac{3}{2}} q_2\right) = \mathrm{const}$.
We take $K$ to vanish as $K(x) \sim \abs{x}^\theta$.
This can then be rearranged as:
\begin{equation}
\label{eq:boundary_equation}
    \partial_x^2 \psi = \frac{-E\psi -K'(x) \partial_x \psi - K_2 \partial_y^2 \psi}{K(x) + K_2}.
\end{equation}

We now distinguish two regimes:
\begin{enumerate}
    \item $K \sim \abs{x}^\theta, ~\theta>1$: $K'$ is continuous everywhere.
    \item $K \sim \abs{x}^\theta, ~\theta<1$: $K'$ is discontinuous at $x = \pm 1$.
\end{enumerate}
For $\theta > 1$, $K'$ and all other terms in \cref{eq:boundary_equation} are continuous.
Hence, $\partial_x^2 \psi$ is continuous across the boundary.
This is the simple case, with essentially no extra conditions imposed on $\psi$.

Now instead for $\theta < 1$, $K'$ is discontinuous over the boundary, whereas $K+K_2$ and $E\psi$ are continuous.
Additionally, the final term involving $\partial_y^2 \psi$ is continuous under the reasonable assumption that there is a Taylor expansion in the $y$-direction.
This is justified as the singularity is due to the crossing in $x$, whereas the position of $y$ along the line is irrelevant.
Combining all this, we have the dominant behavior for $\theta < 1$ being:
\begin{equation}
    \partial_x^2 \psi \sim \frac{-K'(x) \partial_x \psi}{K_2}.
\end{equation}
To extract information on $\partial_x \psi$, we integrate this equation from $0$ to $\epsilon$ and also from $0$ to $-\epsilon$, where $\epsilon$ is some small value.
To one side, where $K(x) = 0$, we have $\partial_x^2 \psi \sim (\mathrm{well~behaved~functions})$, so $\partial_x \psi$ is also well behaved.
To the other side, we integrate to find
\begin{equation}
    \partial_x \psi \rvert_\epsilon \sim (\mathrm{const}) + (\mathrm{const}) \epsilon^\theta.
\end{equation}
These two conditions on either side of the internal line, which qualitatively match with lattice numerics, provide the matching conditions for wavefunctions in the $\theta < 1$ regime.

\section{Symmetries of the problem}
\label{app:hexagon_symmetries}
The symmetries of the three particle problem in original coordinates are generated by three actions: cyclic permutation $R$, exchange $X$ and parity $P$, defined as
\begin{align}
    R: \begin{pmatrix}
        x_1 \\
        x_2 \\
        x_3
    \end{pmatrix}   &\mapsto  \begin{pmatrix}
        x_2 \\
        x_3 \\
        x_1
    \end{pmatrix},~\begin{pmatrix}
        p_1 \\
    p_2 \\
        p_3
    \end{pmatrix}   \mapsto  \begin{pmatrix}
        p_2 \\
        p_3 \\
        p_1
    \end{pmatrix}\\
    X: \begin{pmatrix}
        x_1 \\
        x_2 \\
        x_3
    \end{pmatrix}   &\mapsto  \begin{pmatrix}
        x_2 \\
        x_1 \\
        x_3
    \end{pmatrix},  ~\begin{pmatrix}
        p_1 \\
        p_2 \\
        p_3
    \end{pmatrix}   \mapsto  \begin{pmatrix}
        p_2 \\
        p_1 \\
        p_3
    \end{pmatrix},\\
    P: x_a &\mapsto -x_a,~ p_a \mapsto - p_a
\end{align}
and all combinations of which there are 12 distinct ones, forming the dihedral group $D_{6}$ containing 12 elements. This is precisely the symmetry group of a regular hexagon. This is seen by passing onto reduced coordinates (we will focus on the position coordinates with identical results on momenta)
\begin{equation}
    q_1 = \frac{x_1 - x_2}{\sqrt{2}},~q_2 = \frac{x_1 + x_2 -2 x_3}{\sqrt{6}},~q_3 = \frac{x_1 + x_2 + x_3}{\sqrt{3}}. \nonumber
\end{equation}
The symmetries acting on the reduced coordinates are
\begin{align}
    R: \begin{pmatrix}
        q_1\\
        q_2 \\
        q_3
    \end{pmatrix} &\mapsto \left(
\begin{array}{ccc}
 -\frac{1}{2} & \frac{\sqrt{3}}{2} & 0 \\
 -\frac{\sqrt{3}}{2} & -\frac{1}{2} & 0 \\
 0 & 0 & 1 \\
\end{array} 
\right) \begin{pmatrix}
        q_1\\
        q_2 \\
        q_3
    \end{pmatrix},\\
    X:\begin{pmatrix}
        q_1\\
        q_2 \\
        q_3
    \end{pmatrix} &\mapsto \left(
\begin{array}{ccc}
 -1 & 0 & 0 \\
 0 & 1 & 0 \\
 0 & 0 & 1 \\
\end{array}
\right) \begin{pmatrix}
        q_1\\
        q_2 \\
        q_3
    \end{pmatrix},\\
    P:\begin{pmatrix}
        q_1\\
        q_2 \\
        q_3
    \end{pmatrix} &\mapsto \left(
\begin{array}{ccc}
 -1 & 0 & 0 \\
 0 & -1 & 0 \\
 0 & 0 & -1 \\
\end{array}
\right) \begin{pmatrix}
        q_1\\
        q_2 \\
        q_3
    \end{pmatrix}. \label{eq:q_symmetry}
\end{align}
We see that, fortuitously, the reduced coordinates also block-diagonalize the symmetry actions. The $q_1,q_2$ block forms a 2d irreducible representation (irrep) of $D_{6}$ whereas $q_3$ forms a non-trivial 1d irrep. The action on the $q_1,q_2$ plane are precisely transformations of the hexagon. 

What are the structures of the eigenstates of the quantized 3-fracton Hamiltonian in reduced coordinates? $D_{6}$ has exactly four 1d irreps (including the trivial one), and two 2d irreps. The eigenstates can correspond to any of these six irreps. In other words, the spectrum consists of four kinds of singlets and two kinds of doublets. If 
 $u_g$ are the $2\times 2$ symmetry blocks in \cref{eq:q_symmetry} and the coordinates transform as $ \begin{pmatrix}
        q'_1\\
        q'_2
    \end{pmatrix} = u_g\begin{pmatrix}
        q_1\\
        q_2
    \end{pmatrix}$, the eigenstates transform as
\begin{align}
   \text{Singlets: }& \psi(q'_1,q'_2) = e^{i\theta_g} \psi(q_1,q_2)\\
   \text{Doublets: }& \psi_a(q'_1,q'_2) = w(g)_{ab} \psi_b(q_1,q_2),~
\end{align}
The various possibilities for $\theta_g$ and $w_g$ are listed in \cref{tab:D6}, up to a basis choice.

For a system with $N$ fractons, the system is invariant under arbitrary permutations, i.e.\ under the symmetric group $S_N$.
As a result of this symmetry, the eigenstates can be organized into irreducible representations of $S_N$ (extended by an additional parity symmetry), analogous to $D_6$ for 3 fractons.
If we were to treat the particles as indistinguishable, depending on whether the particles are bosons or fermions,  we only retain the eigenstates corresponding to the one-dimensional signed or unsigned irreducible representations of $S_N$, wherein the wavefunctions pick up a sign $\pm1$ under coordinate exchange.
Therefore, our current framework already contains all the information to treat the particles as bosons or fermions.
Clustering is a property of all symmetry sectors. As such, we would observe clustering for fermionic or bosonic fractons.
However, for a larger number of particles, for example, if there were many particles within a single cluster, we would expect the Fermi degeneracy pressure to play a significant role; as such, bosons and fermions would behave differently in larger systems.

\begin{table}[!t]
    \centering
    \begin{tabular}{c||c|c|c|c|c|c}
       $g$ &        $\theta^A_g$  &      $\theta^B_g$  &       $\theta^C_g$ &       $\theta^D_g$ & $w^A_g$ & $w^B_g$\\
       \hline 
       \hline
        R.P & 1 & -1 & 1 & -1 & $\begin{pmatrix}
            e^{i \pi/3} & 0 \\
            0 & e^{-i\pi/3}
        \end{pmatrix} $   & $\begin{pmatrix}
            0 &1 \\
            1 & 0 
        \end{pmatrix}$ \\
        \hline
        X & 1 & 1 & -1 & -1 & $\begin{pmatrix}
            e^{2i \pi/3} & 0 \\
            0 & e^{-2i\pi/3}
        \end{pmatrix} $   & $\begin{pmatrix}
            0 &1 \\
            1 & 0 
        \end{pmatrix}$
    \end{tabular}
    \caption{Six irreps of $D_{6}$ that the eigenstates can transform as (up to a basis choice). The representations can be fully specified from their action for $R.P$ and $X$ symmetries. }
    \label{tab:D6}
\end{table}

\section{Continuum formulation}
\label{app:continuum}
Recall that when one passes from first to second quantization, we make the following replacements for one body $A_i$ and two body $B_{ij}$ first-quantized operators
\begin{align}
\sum_{i=1}^N A_i &\xrightarrow{\text {II-quant}} \int d^d x \phi^{\dagger}(\vec{x}) \bra{\vec{x}}A \ket{\vec{x}} \phi(\vec{x}),\\
\sum_{i,j=1}^N B_{ij} &\xrightarrow{\text {II-quant}} \int d^d x d^d y \phi^{\dagger}(\vec{x}) \phi^{\dagger}(\vec{y}) \bra{\vec{x},\vec{y}}B \ket{\vec{x},\vec{y}} \phi(\vec{x})\phi(\vec{y}),
\end{align}
We want to apply this to our classical Hamiltonian 
\begin{multline}
	H = \lambda_1 \sum_{j , k=1}^N K_1(|\vec{x}_j-\vec{x}_k|)  \left(\vec{p}_j - \vec{p}_k \right)^2 \\+ \lambda_2\sum_{j , k=1}^N K_2(|\vec{x}_j-\vec{x}_k|) \left(\vec{p}_j - \vec{p}_k \right)^4  \label{eq:H_manybody_app}.
\end{multline}
All terms are two-body operators and thus we expect the second-quantized model to be quartic in field operators. We proceed term-by-term. We pick the symmetric operator ordering as in the main text and work in the position basis:
\begin{widetext}
\begin{multline}
\begin{split}
    \sum_{j , k=1}^N (p^\mu_{j} - p^{\mu}_k) K_1(|\vec{x}_j-\vec{x}_k|)  (p^\mu_{j} - p^{\mu}_k)  \xrightarrow{\text {II-quant}}  
    -\int d^dx d^dy~\phi^\dagger(\vec{x}) \phi^\dagger(\vec{y}) \innerproduct{\vec{x} ,\vec{y}}{(p^\mu_{x} - p^{\mu}_y) K_1(|\vec{x}-\vec{y}|)  (p^\mu_{x} - p^{\mu}_y)|\vec{x} ,\vec{y}}\\
   = -\int d^dx d^dy~\phi^\dagger(\vec{x}) \phi^\dagger(\vec{y}) \left( \frac{\partial}{\partial x^\mu} -  \frac{\partial}{\partial y^\mu} \right) K_1(|\vec{x}-\vec{y}|) \left( \frac{\partial}{\partial x^\mu} -  \frac{\partial}{\partial y^\mu} \right) \phi(\vec{x}) \phi(\vec{y}) \\ \xrightarrow{\text{IBP}}  \int d^dx d^dy K_1(|\vec{x}-\vec{y}|) \left|\phi(\vec{y}) \vec{\nabla}_x \phi(\vec{x})  - \phi(\vec{x}) \vec{\nabla}_y \phi(\vec{y})  \right|^2. \label{eq:H1_2quant_bilocal}
\end{split}
\end{multline}
\end{widetext}
IBP refers to integration by parts in both variables. However, if we take the ultra local limit $K_1(x) \rightarrow \delta(x)$, \cref{eq:H1_2quant_bilocal} vanishes. But if we take a different operator ordering as follows,
\begin{widetext}
\begin{multline}
\begin{split}
    \frac{1}{2}\sum_{j , k=1}^N K_1(|\vec{x}_j-\vec{x}_k|)  (\vec{p}_{j} - \vec{p}_k)^2 + h.c.  \xrightarrow{\text {II-quant}}  
    -\frac{1}{2}\int d^dx d^dy~\phi^\dagger(\vec{x}) \phi^\dagger(\vec{y}) K_1(|\vec{x}-\vec{y}|) \left( \vec{\nabla}_x - \vec{\nabla}_y \right)^2 \phi(\vec{x}) \phi(\vec{y}) \\ = - \frac{1}{2}\int d^dx d^dy~ K_1(|\vec{x}-\vec{y}|)  \phi^\dagger(\vec{x}) \phi^\dagger(\vec{y})\left(\phi(\vec{y}) {\nabla}^2_x \phi(\vec{x}) +\phi(\vec{x}) {\nabla}^2_y \phi(\vec{y})  + \vec{\nabla}_x\phi(\vec{x}) .\vec{\nabla}_y \phi(\vec{y})  \right) + h.c. \label{eq:H1_2quant_bilocal_good}
\end{split}
\end{multline}
\end{widetext}
the ultra-local limit of \cref{eq:H1_2quant_bilocal} does not vanish and we get
\begin{equation}
    - \frac{1}{2}\int d^dx (\phi^\dagger)^2 \left(\phi\vec{\nabla}^2 \phi -  \vec{\nabla} \phi. \vec{\nabla} \phi\right)+ h.c. \label{eq:H1_ultralocal}
\end{equation}
Similar steps also give us the expression for the second term. Using the symmetric ordering again, we have
\begin{widetext}
\begin{align}
\begin{split}
    \sum_{j , k=1}^N (p^\alpha_{j} - p^{\alpha}_k) (p^\beta_{j} - p^{\beta}_k)& K_2(|\vec{x}_j-\vec{x}_k|) (p^\alpha_{j} - p^{\alpha}_k) (p^\beta_{j} - p^{\beta}_k)\xrightarrow{\text {II-quant}}  \\& \int d^dx d^dy K_2(|\vec{x}-\vec{y}|) \left|\phi(\vec{x}) \frac{\partial^2 \phi(\vec{y})}{\partial y_\alpha \partial y_\beta}  +\phi(\vec{y}) \frac{\partial^2 \phi(\vec{x})}{\partial x_\alpha \partial x_\beta} - \frac{\partial \phi(\vec{x})}{ \partial x_\alpha} \frac{\partial \phi(\vec{y})}{ \partial y_\beta}  - \frac{\partial \phi(\vec{x})}{ \partial x_\beta} \frac{\partial \phi(\vec{y})}{ \partial y_\alpha} \right|^2. \label{eq:H2_2quant_bilocal}
\end{split}
\end{align}
\end{widetext}
The ultra-local limit $K_{2}(x) \rightarrow \delta(x)$ can directly be applied to \cref{eq:H1_2quant_bilocal} which does not vanish but gives 
\begin{multline}
    4 \int d^dx \lambda_2 \left|\phi \partial_\alpha \partial_\beta \phi - \partial_\alpha \phi \partial_\beta \phi\right|^2 \label{eq:H2_ultralocal}
\end{multline}
Using \cref{eq:H1_ultralocal,eq:H2_ultralocal} gives us the second quantized version of \cref{eq:H_manybody_app} in the ultra-local limit, leading to the form in the main text~\cref{eq:Pretko}:
\begin{multline}
       \mathcal{H} = -\lambda_1 (\phi^\dagger)^2 \left(\phi\vec{\nabla}^2 \phi -  \vec{\nabla} \phi. \vec{\nabla} \phi\right)+ h.c.\\
    +4 \lambda_2 \left|\phi \partial_\alpha \partial_\beta \phi - \partial_\alpha \phi \partial_\beta \phi\right|^2. \label{eq:Pretko_app}
\end{multline}

\bibliography{references}

@misc{GromovRadzihovsky2022fractonReview,
      title={Fracton Matter}, 
      author={Gromov, Andrey and Radzihovsky, Leo},
      year={2022},
      eprint={2211.05130},
      archivePrefix={arXiv},
      primaryClass={cond-mat.str-el}
}

@misc{sadki2025phasespacefractons,
      title={Phase space fractons}, 
      author={Ylias Sadki and Abhishodh Prakash and S. L. Sondhi and Daniel P. Arovas},
      year={2025},
      eprint={2502.02650},
      archivePrefix={arXiv},
      primaryClass={cond-mat.stat-mech}
}

@Inbook{ChalkerOBD2011,
author="Chalker, John T.",
title="Geometrically Frustrated Antiferromagnets: Statistical Mechanics and Dynamics",
bookTitle="Introduction to Frustrated Magnetism: Materials, Experiments, Theory",
year="2011",
publisher="Springer Berlin Heidelberg",
address="Berlin, Heidelberg",
pages="3--22",
isbn="978-3-642-10589-0",
doi="10.1007/978-3-642-10589-0_1",
url="https://doi.org/10.1007/978-3-642-10589-0_1"
}

@article{MoessnerChalkerOBDPhysRevLett.80.2929,
  title = {Properties of a Classical Spin Liquid: The Heisenberg Pyrochlore Antiferromagnet},
  author = {Moessner, R. and Chalker, J. T.},
  journal = {Phys. Rev. Lett.},
  volume = {80},
  issue = {13},
  pages = {2929--2932},
  numpages = {0},
  year = {1998},
  month = {Mar},
  publisher = {American Physical Society},
  doi = {10.1103/PhysRevLett.80.2929},
  url = {https://link.aps.org/doi/10.1103/PhysRevLett.80.2929}
}

@article{YouMoessner_UVIR_PhysRevB.106.115145,
	title = {Fractonic plaquette-dimer liquid beyond renormalization},
	author = {You, Yizhi and Moessner, Roderich},
	journal = {Phys. Rev. B},
	volume = {106},
	issue = {11},
	pages = {115145},
	numpages = {11},
	year = {2022},
	month = {Sep},
	publisher = {American Physical Society},
	doi = {10.1103/PhysRevB.106.115145},
	url = {https://link.aps.org/doi/10.1103/PhysRevB.106.115145}
}

@article{PretkoChenYou_2020fracton,
	title={Fracton phases of matter},
	author={Pretko, Michael and Chen, Xie and You, Yizhi},
	journal={International Journal of Modern Physics A},
	volume={35},
	number={06},
	pages={2030003},
	year={2020},
	publisher={World Scientific}
}

@article{NandkishoreHermeleFractonsannurev-conmatphys-031218-013604,
	author = {Nandkishore, Rahul M. and Hermele, Michael},
	title = {Fractons},
	journal = {Annual Review of Condensed Matter Physics},
	volume = {10},
	number = {1},
	pages = {295-313},
	year = {2019},
	doi = {10.1146/annurev-conmatphys-031218-013604},

	URL = {

	https://doi.org/10.1146/annurev-conmatphys-031218-013604



	},
	eprint = {

	https://doi.org/10.1146/annurev-conmatphys-031218-013604



	}
}

@article{Chamon_Fracton_PhysRevLett.94.040402,
	title = {Quantum Glassiness in Strongly Correlated Clean Systems: An Example of Topological Overprotection},
	author = {Chamon, Claudio},
	journal = {Phys. Rev. Lett.},
	volume = {94},
	issue = {4},
	pages = {040402},
	numpages = {4},
	year = {2005},
	month = {Jan},
	publisher = {American Physical Society},
	doi = {10.1103/PhysRevLett.94.040402},
	url = {https://link.aps.org/doi/10.1103/PhysRevLett.94.040402}
}

@article{Haah_FractonPhysRevA.83.042330,
	title = {Local stabilizer codes in three dimensions without string logical operators},
	author = {Haah, Jeongwan},
	journal = {Phys. Rev. A},
	volume = {83},
	issue = {4},
	pages = {042330},
	numpages = {16},
	year = {2011},
	month = {Apr},
	publisher = {American Physical Society},
	doi = {10.1103/PhysRevA.83.042330},
	url = {https://link.aps.org/doi/10.1103/PhysRevA.83.042330}
}

@article{VijayHaahFu_FractonDuality_PhysRevB.94.235157,
	title = {Fracton topological order, generalized lattice gauge theory, and duality},
	author = {Vijay, Sagar and Haah, Jeongwan and Fu, Liang},
	journal = {Phys. Rev. B},
	volume = {94},
	issue = {23},
	pages = {235157},
	numpages = {9},
	year = {2016},
	month = {Dec},
	publisher = {American Physical Society},
	doi = {10.1103/PhysRevB.94.235157},
	url = {https://link.aps.org/doi/10.1103/PhysRevB.94.235157}
}

@article{MorningstayKhemaniHust_Thermalization_PhysRevB.101.214205,
  title = {Kinetically constrained freezing transition in a dipole-conserving system},
  author = {Morningstar, Alan and Khemani, Vedika and Huse, David A.},
  journal = {Phys. Rev. B},
  volume = {101},
  issue = {21},
  pages = {214205},
  numpages = {10},
  year = {2020},
  month = {Jun},
  publisher = {American Physical Society},
  doi = {10.1103/PhysRevB.101.214205},
  url = {https://link.aps.org/doi/10.1103/PhysRevB.101.214205}
}

@article{SkinnerPozderac_Thermalization_2023,
	title = {Exact solution for the filling-induced thermalization transition in a one-dimensional fracton system},
	author = {Pozderac, Calvin and Speck, Steven and Feng, Xiaozhou and Huse, David A. and Skinner, Brian},
	journal = {Phys. Rev. B},
	volume = {107},
	issue = {4},
	pages = {045137},
	numpages = {15},
	year = {2023},
	month = {Jan},
	publisher = {American Physical Society},
	doi = {10.1103/PhysRevB.107.045137},
	url = {https://link.aps.org/doi/10.1103/PhysRevB.107.045137}
}

@article{Pretko_FractonGauge_PhysRevB.98.115134,
	title = {The fracton gauge principle},
	author = {Pretko, Michael},
	journal = {Phys. Rev. B},
	volume = {98},
	issue = {11},
	pages = {115134},
	numpages = {6},
	year = {2018},
	month = {Sep},
	publisher = {American Physical Society},
	doi = {10.1103/PhysRevB.98.115134},
	url = {https://link.aps.org/doi/10.1103/PhysRevB.98.115134}
}

@article{GorantlaLamSeiberg_UVIR_PhysRevB.104.235116,
	title = {Low-energy limit of some exotic lattice theories and UV/IR mixing},
	author = {Gorantla, Pranay and Lam, Ho Tat and Seiberg, Nathan and Shao, Shu-Heng},
	journal = {Phys. Rev. B},
	volume = {104},
	issue = {23},
	pages = {235116},
	numpages = {26},
	year = {2021},
	month = {Dec},
	publisher = {American Physical Society},
	doi = {10.1103/PhysRevB.104.235116},
	url = {https://link.aps.org/doi/10.1103/PhysRevB.104.235116}
}

@article{Gromov_Multipole_PhysRevX.9.031035,
	title = {Towards Classification of Fracton Phases: The Multipole Algebra},
	author = {Gromov, Andrey},
	journal = {Phys. Rev. X},
	volume = {9},
	issue = {3},
	pages = {031035},
	numpages = {19},
	year = {2019},
	month = {Aug},
	publisher = {American Physical Society},
	doi = {10.1103/PhysRevX.9.031035},
	url = {https://link.aps.org/doi/10.1103/PhysRevX.9.031035}
}

@article{SalaRakovskyVerresenKnapPollmann_FragmentationPhysRevX.10.011047,
	title = {Ergodicity Breaking Arising from Hilbert Space Fragmentation in Dipole-Conserving Hamiltonians},
	author = {Sala, Pablo and Rakovszky, Tibor and Verresen, Ruben and Knap, Michael and Pollmann, Frank},
	journal = {Phys. Rev. X},
	volume = {10},
	issue = {1},
	pages = {011047},
	numpages = {19},
	year = {2020},
	month = {Feb},
	publisher = {American Physical Society},
	doi = {10.1103/PhysRevX.10.011047},
	url = {https://link.aps.org/doi/10.1103/PhysRevX.10.011047}
}

@article{KhemaniHermeleNandkishore_Shattering_PhysRevB.101.174204,
	title = {Localization from Hilbert space shattering: From theory to physical realizations},
	author = {Khemani, Vedika and Hermele, Michael and Nandkishore, Rahul},
	journal = {Phys. Rev. B},
	volume = {101},
	issue = {17},
	pages = {174204},
	numpages = {17},
	year = {2020},
	month = {May},
	publisher = {American Physical Society},
	doi = {10.1103/PhysRevB.101.174204},
	url = {https://link.aps.org/doi/10.1103/PhysRevB.101.174204}
}

@article{Gromovetal_FractonHydro_PhysRevResearch.2.033124,
  title = {Fracton hydrodynamics},
  author = {Gromov, Andrey and Lucas, Andrew and Nandkishore, Rahul M.},
  journal = {Phys. Rev. Res.},
  volume = {2},
  issue = {3},
  pages = {033124},
  numpages = {11},
  year = {2020},
  month = {Jul},
  publisher = {American Physical Society},
  doi = {10.1103/PhysRevResearch.2.033124},
  url = {https://link.aps.org/doi/10.1103/PhysRevResearch.2.033124}
}

@article{AP2023NRFractons,
  title={Classical nonrelativistic fractons},
  author={Prakash, Abhishodh and Goriely, Alain and Sondhi, SL},
  journal = {Phys. Rev. B},
  volume={109},
  number={5},
  pages={054313},
  year={2024},
  publisher = {American Physical Society},
  doi = {10.1103/PhysRevB.109.054313}
}

@article{classenhowes2025universalfreezingtransitionsdipoleconserving,
  title = {Universal freezing transitions of dipole-conserving chains},
  author = {Classen-Howes, Jonathan and Senese, Riccardo and Prakash, Abhishodh},
  journal = {Phys. Rev. B},
  volume = {112},
  issue = {12},
  pages = {125148},
  numpages = {45},
  year = {2025},
  month = {Sep},
  publisher = {American Physical Society},
  doi = {10.1103/2h1v-yx5l},
  url = {https://link.aps.org/doi/10.1103/2h1v-yx5l}
}

@article{babbar2025classicalfractonslocalchaos,
  title={Classical fractons: Local chaos, global broken ergodicity, and an arrow of time},
  author={Babbar, Aryaman and Sadki, Ylias and Prakash, Abhishodh and Sondhi, SL},
  journal={Physical Review B},
  volume={111},
  number={24},
  pages={245134},
  year={2025},
  publisher={APS},
  doi = {10.1103/g2l1-s2vy}
}

@article{AP2024MachianFractons,
  title={Machian fractons, Hamiltonian attractors, and nonequilibrium steady states},
  author={Prakash, Abhishodh and Sadki, Ylias and Sondhi, SL},
  journal={Physical Review B},
  volume={110},
  number={2},
  pages={024305},
  year={2024},
  publisher={APS},
  doi = {10.1103/PhysRevB.110.024305}
}

@book{zettl2005sturm,
  title={Sturm-Liouville Theory},
  author={Zettl, Anton},
  number={121},
  year={2005},
  publisher={American Mathematical Soc.}
}

@article{littlejohn2011legendre,
  title={The Legendre equation and its self-adjoint operators},
  author={Littlejohn, Lance L and Zettl, Anton},
  journal={Electronic Journal of Differential Equations},
  volume={2011},
  number={69},
  pages={1--33},
  year={2011}
}

@article{stuart2018stability,
  title={STABILITY ANALYSIS FOR A FAMILY OF DEGENERATE SEMILINEAR PARABOLIC PROBLEMS.},
  author={Stuart, Charles A},
  journal={Discrete \& Continuous Dynamical Systems: Series A},
  volume={38},
  number={10},
  year={2018}
}

@book{reed_simon1,
  title={Methods of modern mathematical physics},
  author={Reed, Michael and Simon, Barry},
  volume={1},
  year={1972},
  publisher={Academic press New York}
}

@article{berry1972semiclassical,
  title={Semiclassical approximations in wave mechanics},
  author={Berry, Michael V and Mount, KE},
  journal={Reports on Progress in Physics},
  volume={35},
  number={1},
  pages={315},
  year={1972},
  publisher={IOP Publishing}
}

@article{atkinson1960wave,
  title={Wave propagation and the Bremmer Series},
  author={Atkinson, FV},
  journal={Journal of mathematical analysis and applications},
  volume={1},
  number={3-4},
  pages={255--276},
  year={1960},
  publisher={Elsevier}
}

@book{hille1969lectures,
  title     = {Lectures on Ordinary Differential Equations},
  author    = {Hille, Einar},
  year      = {1969},
  publisher = {Addison-Wesley}
}

@book{fetter2012quantum,
  title={Quantum theory of many-particle systems},
  author={Fetter, Alexander L and Walecka, John Dirk},
  year={2012},
  publisher={Courier Corporation}
}

@book{heading2013introduction,
  title={An introduction to phase-integral methods},
  author={Heading, John},
  year={2013},
  publisher={Courier Corporation}
}

@article{shender1982antiferromagnetic,
  title={Antiferromagnetic garnets with fluctuationally interacting sublattices},
  author={Shender, EF},
  journal={Soviet Journal of Experimental and Theoretical Physics},
  volume={56},
  number={1},
  pages={178},
  year={1982}
}

@article{henley1987ordering,
  title={Ordering by disorder: Ground-state selection in fcc vector antiferromagnets},
  author={Henley, Christopher L},
  journal={Journal of Applied Physics},
  volume={61},
  number={8},
  pages={3962--3964},
  year={1987}
}

@article{pretko2017emergent,
  title={Emergent gravity of fractons: Mach’s principle revisited},
  author={Pretko, Michael},
  journal={Physical Review D},
  volume={96},
  number={2},
  pages={024051},
  year={2017},
  publisher={APS}
}

@article{Yuan2020FractonicSF,
  author       = {Jian-Keng Yuan and Shuai A. Chen and Peng Ye},
  title        = {Fractonic superfluids},
  journal      = {Phys. Rev. Res.},
  volume       = {2},
  pages        = {023267},
  year         = {2020},
  doi          = {10.1103/PhysRevResearch.2.023267}
}

@article{Chen2021FractonicSF2,
  author       = {Shuai A. Chen and Jian-Keng Yuan and Peng Ye},
  title        = {Fractonic superfluids. II. Condensing subdimensional particles},
  journal      = {Phys. Rev. Res.},
  volume       = {3},
  pages        = {013226},
  year         = {2021},
  doi          = {10.1103/PhysRevResearch.3.013226}
}

@article{Lake2022DipolarBH,
  author       = {Ethan Lake and Michael Hermele and T. Senthil},
  title        = {Dipolar Bose–Hubbard model},
  journal      = {Phys. Rev. B},
  volume       = {106},
  pages        = {064511},
  year         = {2022},
  doi          = {10.1103/PhysRevB.106.064511}
}

@article{Lake2022NonFermi,
  author       = {Ethan Lake and T. Senthil},
  title        = {Non-Fermi liquids from kinetic constraints in tilted optical lattices},
  journal      = {Phys. Rev. Lett.},
  volume       = {131},
  pages        = {043403},
  year         = {2023},
  doi          = {10.1103/PhysRevLett.131.043403}
}

\end{document}